\documentclass[preprint,prc,aps,tightenlines,showpacs,floatfix]{revtex4-1}

\usepackage{graphicx}
\usepackage{bm}
\usepackage{amsmath}
\usepackage{amssymb}

\newcommand{\sfrac}[2] {\ensuremath{\textstyle\frac{#1}{#2}}}

\begin{document}

\title{Very low-energy nucleon-$^{16}$O  coupled-channel scattering:
results with a phenomenological vibrational model.}

\author{J. P. Svenne$^{(1)}$}
\email{svenne@physics.umanitoba.ca}
\author{L. Canton$^{(2)}$}
\email{luciano.canton@pd.infn.it} 
\email{Corresponding author}
\author{K. Amos$^{(3,5)}$}
\email{amos@unimelb.edu.au}
\author{P. R. Fraser$^{(4)}$}
\email{paul.fraser@curtin.edu.au}
\author{\linebreak S. Karataglidis$^{(5,3)}$}
\email{stevenka@uj.ac.za}
\author{G. Pisent$^{(6)}$}
\email{w.pisent@gmail.com}
\author{D. van der Knijff$^{(3)}$}
\email{dvanderknijff@gmail.com}

\affiliation{$^{(1)}$ Department  of  Physics  and Astronomy,
    University of Manitoba, and Winnipeg Institute for Theoretical Physics,
    Winnipeg, Manitoba, Canada R3T 2N2}
\affiliation{$^{(2)}$ Istituto  Nazionale  di  Fisica  Nucleare,
    Sezione di Padova, Padova I-35131, Italia}
\affiliation{$^{(3)}$ School  of Physics,  University of  Melbourne,
    Victoria 3010, Australia}
\affiliation{$^{(4)}$ Department of Physics, Astronomy and Medical Radiation Sciences,
 Curtin University, GPO Box U1987, Perth 6845,  Australia}
\affiliation{$^{(5)}$ Department of Physics, University of
  Johannesburg,   P.O. Box 524 Auckland Park, 2006, South Africa}
\affiliation{$^{(6)}$ Dipartimento di Fisica e Astronomia, Universit\`{a} di Padova,
Padova, I-35131, Italia}

\date{\today}

\begin{abstract}

We employ a collective vibration coupled-channel model to describe the 
nucleon-${}^{16}$O cluster systems, obtaining low-excitation spectra for ${}^{17}$O and
${}^{17}$F. Bound and resonance states of the compound systems have been deduced, showing good
agreement with experimental spectra. Low-energy scattering cross sections of neutrons and protons 
from ${}^{16}$O also have been calculated and the results compare well with available experimental
data.

\end{abstract}


\maketitle

\section{Introduction}
\label{sec1}

We apply the multi-channel algebraic scattering method (MCAS) to study the bound and resonance properties of the 
$^{17}$O and $^{17}$F nuclei below and above the nucleon-core threshold. With the same method, we investigate nucleon elastic scattering on $^{16}$O 
at very low energies. We introduce as essential physics ingredients the coupling of the incident, or valence, nucleon with the low-lying collective vibrational states of the $^{16}$O core. In the past, we considered many applications of the MCAS method describing couplings of valence nucleons with rotational states of the core. This is the first exploratory study where we consider collective vibrations in the MCAS method.

The approach we propose herein has some similarities, and to a certain extent is complementary, to the microscopic particle-vibration coupling (PVC) method developed recently to calculate scattering cross-sections in light-medium nuclei. In the PVC method the collective core excitations are treated microscopically with the RPA method, and one can find significant developments and applications along these lines in Refs.~\cite{Ka12, Bl15, Nh15, Mi12, No10}. These microscopic type calculations are quite promising in the description of nucleon-nucleus collisions in the moderately low-energy range between 10 and 40 MeV, particularly in describing the particle-hole states as a doorway-state mechanism through which the flux evolves into more complex configurations such as overlapping states of the compound 
nucleus ~\cite{No10, Ka12}. All these approaches, while successful at moderate 
energies, do not describe adequately the cross section in the very low-energy regimes 
(typically, lower than 5 MeV).

To consider nucleon scattering on $^{16}$O in the very low-energy regime, we introduce 
a purely phenomenological description of excited states in terms of collective vibrations 
of the core nucleus. We use a geometrical model for particle-vibration couplings 
as has been discussed in textbooks \cite{Bo75}. We do not attempt to define the connection with the microscopic origin of the ingredients we use, but simply employ the basic Hamiltonian coupled-channel parameters in a fitting procedure.
In addition, we consider couplings generated only by quadrupole and octupole phonons, which dominate the low-energy regime. These limitations, however, are not inherent to the MCAS approach. In the future it is feasible to apply the method to particle-vibration couplings generated microscopically, or to include additional multipolarities (for example direct dipole or monopole excitation modes, etc.) that presently are not taken into account.

Previously, the MCAS method was developed and first applied~\cite{Am03} to the well-studied $n + ^{12}$C system. That first MCAS investigation focussed on obtaining excellent agreement with the experimental total elastic $n + ^{12}$C cross section to $\sim 4$ MeV, by varying free parameters of the potential used. 
With these same parameters, the spectrum of $^{13}$C to $\sim$ 8 MeV also was well described, for both bound states and resonances in the compound nucleus,  $^{13}$C. A number of MCAS studies on other nucleon plus nucleus systems 
have been carried out and published since then, see, e.g.~\cite{Ca05,Ca06, Ca06a,Am12}. 
In all of those studies, a rotational model was used to specify the matrix of coupled-channel
interaction potentials for the nucleon with each target nucleus.

 $^{12}$C is a partially 
closed-shell nucleus in which the $0s_\frac{1}{2}$ state is expected to be fully occupied, 
the $0p_\frac{3}{2}$ and $0p_\frac{1}{2}$ states partially occupied.  
Conversely, the structures of $^{17}$O and $^{17}$F have been assumed to be that of a nucleon coupled to the $^{16}$O core, with the latter defined as a closed $0p$ shell nucleus. That model yields the single nucleon energies in the $0d1s$ shell model, which are obtained for the positive parity states of both $^{17}$O and $^{17}$F. Yet that model is too simplistic: the prevalence of low-lying negative parity states in both mass-17 nuclei is a consequence of the $^{16}$O core being far more complex: Brown and Green~\cite{Br66} had first realised that at the minimum a $4\hbar\omega$ shell model is needed to describe the
spectrum of $^{16}$O. Haxton and Johnston~\cite{Ha90} had performed such a large-scale calculation, which was able to reproduce the positive parity states of $^{16}$O, especially the first excited state, which is the $0^+_2$ state at 6.06~MeV. Negative parity states were calculated in Ref.~\cite{Ka96}, using the same interaction and single-particle model space in a restricted $(1+3+5)\hbar\omega$ model space calculation. This idea is expanded in Appendix A, where a comparison is made of the spectrum obtained from the shell model and from MCAS.

In treating the coupling of a nucleon to the $^{16}$O core, one must encompass the complicated multi-$\hbar\omega$ description of the core by a coupled channels description, for which a vibrational model description of the states in $^{16}$O is appropriate. Therefore, in Appendix B, we discuss in detail the particle-vibration coupling interactions to be used in the coupled-channel model.



As has been demonstrated in~\cite{Ca05}, solutions of the coupled-channel problem could have some spuriosity due to violation of the Pauli principle by single-nucleon orbit occupancies, in attaching the valence nucleon to states  of the core that are already filled.
However, it is possible to ensure that the Pauli principle is obeyed in coupled-channel 
problems 
by using orthogonalizing pseudo-potentials 
(OPP)~\cite{Ca05,Ca06a}. Detailed clarification of this procedure 
is given in~\cite{Ca05,Am13}.  In the first calculations~\cite{Am03} of the  nuclear system studied 
with MCAS, ($n$+${}^{12}$C), Pauli blocking was required for the $0s_{\frac{1}{2}}$ 
and $0p_{\frac{3}{2}}$ neutron orbits. That study allowed all other orbits in the target states 
used to be accessible in the cluster solutions. More specifics are given in \cite{Am03}
and articles published subsequently~\cite{Ca05,Ca06a,Am13}.
 
To construct the couplings of a valence nucleon (or projectile) with a
nucleus by using a geometric collective model, the coupling interactions
are classified by coupling parameters, $\beta_L$. These parameters are required
in a coupled-channel Hamiltonian for the coupling of the valence nucleon
with the low-excitation states of the core nucleus. These $\beta_L$ therefore,
are not necessarily given from EM transition data of the core (because
they involve also the interaction effects with the extra nucleon), but
could be comparable to such. MCAS vibrational model results for the
nucleon-${}^{16}$O clusters are given in Sections~\ref{sec2} and ~\ref{subC}, where, in the
former, coupling strengths $\beta_L$ are allowed to be free parameters,
and in the latter they are associated with deformations known from other data analyses.
In Section ~\ref{subEM} we summarize the EM properties for $^{16}$O to be expected with the described collective model.

Section~\ref{sec6} contains the conclusions.


\section{MCAS results for the $n$+${}^{16}$O and $p$+${}^{16}$O systems}
\label{sec2}

 We use as the primary nuclear interaction Hamiltonian (between the 
odd nucleon and the core) the following potential form,
\begin{equation}
V(r) = 
\left[ V_0 + V_{ll} {\{\bf l \cdot l}\} + V_{ss} {\{\bf I \cdot s}\} 
\right] w(r) + 2\lambda_\pi^2 
V_{ls} \frac{1}{r} \frac{\partial w(r)}{\partial r} {\{\bf l \cdot s}\}\ .
\label{poteq-FIRST}
\end{equation}
A Woods-Saxon shape, $w(r) = \left[1 + \exp\left(\frac{r-R_0}{a} \right) \right]^{-1}$, 
has been used. The vector operators ${\bf l, s, I}$ denote orbital, nucleon spin, and target spin, respectively.

The interaction contains operator components with zero, first, and second order
irreducible terms due to the expansion of the vibration/deformation operator. 
For each term in the
interaction, the coupled-channel expressions in the channel-coupling scheme can be given as
\begin{align}
V_{cc'}(r) = \left\{V^{(0)}(r)\right\}_{cc'} 
&+\;  \left\{V^{(1)}(r) \sum_\lambda 
{\mathbf {\cal Q}_\lambda^{(1)} \cdot Y_\lambda}(\theta \phi) \right\}_{cc'} 
\nonumber\\
&\hspace*{0.5cm}+\; \left\{V^{(2)}(r) \sum_\lambda 
\left[ \sum_{l_1 l_2} {\cal Q}_\lambda^{(2)}(l_1, l_2)\right] 
\cdot Y_\lambda(\theta \phi) \right\}_{cc'}\ .
\label{Mcaspot1-FIRST}
\end{align}
The approach is explained in full detail in Appendix B.
The importance of including second-order terms in the deformation expansion of the interaction has been discussed in Ref.~\cite{AMOS05}.


In coordinate space, if those potentials are designated by local forms $V_{cc'}(r)
  \delta(r-r')$, the application of OPP method requires considering the solutions of the Schr\oe dinger equation 
with the generalized nonlocal  potential
  \begin{equation}
    \mathcal{V}_{cc'}(r,r')  =  V_{cc'}(r)\delta(r-r')  +  \lambda_c
    A_c(r) A_c(r')\delta_{cc'} ,
  \end{equation}
  where $A(r)$ is  the radial part of the  single-particle bound-state
  wave function  in channel $c$  spanning the phase space  excluded by
  the Pauli principle. The OPP method takes into account the Pauli forbidden states
 in the limit $\lambda_c \to \infty$, and for pratical use $\lambda_c = 10^6$~MeV suffices.
But we take into account also more general configurations with smaller values for $\lambda_c$
as extensively discussed in Refs.~\cite{Ca06a,Am13}.

The full set of parameters that defines the interaction potential is given in Table~\ref{param-tab}.

\begin{table}[ht]
\begin{ruledtabular}
\caption{\label{param-tab}
Parameter values for $n+^{16}$O and $p+^{16}$O MCAS cluster structure.
 The potential parameters, $V_0$, $V_{\ell \ell}$, $V_{\ell s}$, $V_{s s}$, have different values, if they act on negative or positive orbital parity states, $P = -$ or $P = +$, respectively. The lower part of the Table describes the $\lambda_c$ parameters of the OPP term.}
\begin{tabular}{cccccc}
$V_x$ (MeV) & $P = -$ & $P = +$ & Geometry & value &  Coulomb~\cite{Vr87}\\
\hline
$V_0$ & $-$47.15 & $-$50.6 & $R_0$ & 3.15 fm & $R_c$ = 2.608 fm \\
$V_{\ell \ell}$ & 2.55 & 0.0 & a & 0.65 fm & $a_c$ = 0.513 fm\\
$V_{\ell s}$ & 6.9 & 7.2 & $\beta_2$ &  0.21 &  $w$ = $-$0.051\\
$V_{s s}$ & 2.5 & $-$2.0 & $\beta_3$ & 0.42 & \\
\hline
$I^{\pi}_n$ & $E_n$ (MeV) & $0s_{\frac{1}{2}}$ & $0p_{\frac{3}{2}}$ & $0p_{\frac{1}{2}}$ & $0d_{\frac{5}{2}}$ \\
\hline
$0^+_1$ & 0.0 & $10^6$ & $10^6$ & $10^6$ & 0.0 \\
$0^+_2$ & 6.049 & $10^6$ & $10^6$ & 0.0 & 0.0 \\
$3^-_1$ & 6.13 & $10^6$ & $10^6$ & 5.0 & 0.0 \\
$2^+_1$ & 6.92 & $10^6$  & $10^6$ & 0.0 & 0.0 \\
$1^-_1$ & 7.12 & $10^6$  & $10^6$  & 5.0 & 1.0 \\
\end{tabular}
\end{ruledtabular}
\end{table}

MCAS calculations were carried out for $n$+$^{16}$O, 
using 5 target states in $^{16}$O, namely, the $0^+$ ground state ($E = 0$ MeV), 
the second $0^+$ state ($E = 6.049$ MeV), the first $3^-$ state ($E = 6.1299$ MeV), 
the first $2^+$ state ($E = 6.9171$ MeV), and the first $1^-$ state ($E = 7.1169$ MeV). 
These states, along with the corresponding Pauli blocking or hindrance strengths
are listed in the lower section of Table~\ref{param-tab}. 
The use of blocking strengths with dimensions of energy is 
typical of approaches that use the OPP method; a method that is not restricted only 
to nuclear physics applications. It was applied also in studies of electronic structure of atoms
to eliminate unwanted states in bound~\cite{Mi99} and scattering~\cite{Iv03}  problems. In the Table,
the blocking strengths are given in MeV.

The $2_1^+$ and $3_1^-$ states are considered to be single-phonon states.
In the present model, we include also the couplings of the excited  $0_2^+$ and $1_1^-$ states to the $0_1^+$ ground state,  
but only as a second-order effect in the couplings parameters (see Appendix B). We did not include couplings with direct excitation modes
described by monopole or dipole operators. In this sense the model approach used here is more schematic 
than the microscopic approach used in Ref.~\cite{giancolo}. However that microscopic approach was designed to describe excitations of giant-type resonances which are located at higher energies.

All the other potential and geometric parameters used for the present nucleon$+^{16}$O calculations are given 
in the upper part of Table~\ref{param-tab}.
We note that the value of $\beta_2$ in the table is small compared to
values required in assessment of a  B(E2) value in ${}^{16}$O~\cite{SR01}
and later, in the next section, we consider the effects of setting the 
deformation parameter in MCAS evaluations to match the electromagnetically 
determined value.

\begin{figure}[ht]
\scalebox{0.75}{\includegraphics{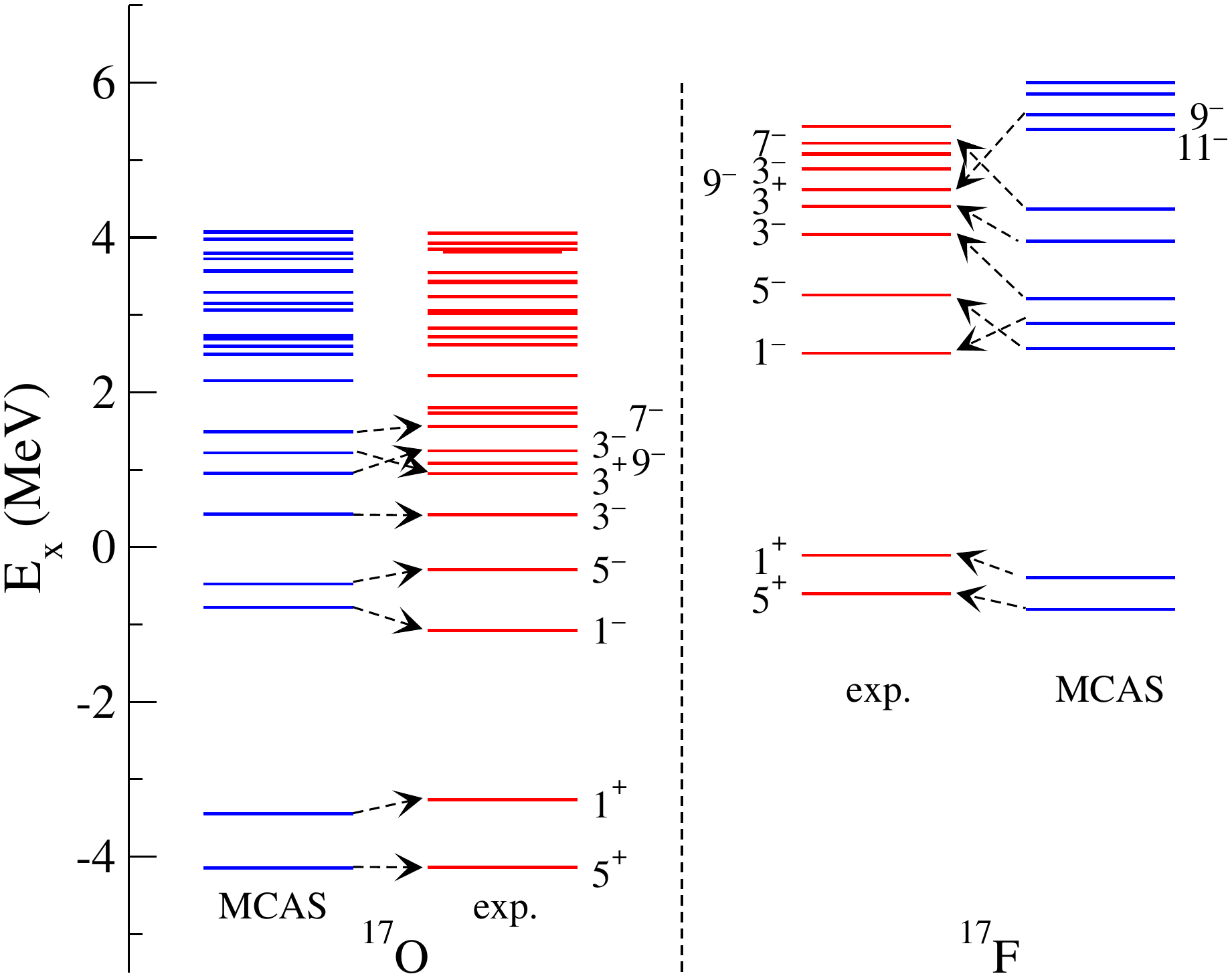}}
\caption{\label{Fig3} (color online)
Spectra of $^{17}$O,  (left panel) calculated with MCAS  and  experimental; 
and of $^{17}$F (right panel), using the parameter set of Table~\ref{param-tab}.
The numbers by the levels are twice the spin of the level, and the superscript  indicates 
the parity.  The zero on the scale is at the respective 
$n + ^{16}$O and $p + ^{16}$O thresholds.}
\end{figure}

The MCAS results found with this parameter set of Table~\ref{param-tab} are compared to the known  spectra of $^{17}$O
(left columns) and of $^{17}$F (right columns) in Fig.~\ref{Fig3}. 
There are more evaluated levels found at higher energies than shown. 
Good agreement between theory and data at low energies is now obtained for both nuclear systems.

\subsection{Bound and resonant states in  $^{17}$O and $^{17}$F.}
\label{subA}

In  Table~\ref{Low30-NEW}, we give the spin/parity values, the data, and  MCAS results 
for $^{17}$O, and the data and MCAS results for $^{17}$F for the lowest 30 levels of each.
The oxygen list is sorted in increasing energy of the experimental values. 
In Table~\ref{Low30-NEW}, the column for the measured $^{17}$F levels are always associated 
with a $J^{\pi}$ value, even though some measured values have unknown $J^{\pi}$.
In those cases where the association of $J^{\pi}$ is purely speculative, 
the energy levels are endowed with an asterisk.
Note that the experimental widths are full widths at half maximum, and are total widths, 
while the MCAS widths, $\Gamma_{mcas}$, represent only nucleon-emission widths.
\begin{table}[t]
\begin{ruledtabular}
\caption{\label{Low30-NEW}
The 30 lowest levels in $^{17}$O and $^{17}$F, experiment and theory.  Energy levels are in MeV,
widths in keV.}
\begin{tabular}{ccccccccc}
\hline
$J^{\pi}$ & $^{17}$O:  $E_{exp}$ & $\Gamma_{exp}$ &   $^{17}$O:  $E_{mcas}$ & $\Gamma_{mcas}$ &
$^{17}$F:  $E_{exp}$ & $\Gamma_{exp}$ &  $^{17}$F:  $E _{mcas}$ & $\Gamma_{mcas}$ \\
\hline
${\frac{5}{2}}^+$  & -4.1436 & -- & -4.1432 & -- & -0.6005 & -- & -0.8079 & -- \\
${\frac{1}{2}}^+$  & -3.27287 & -- & -3.4426 & --  & -0.10517 & -- & -0.3927 & -- \\
${\frac{1}{2}}^-$  & -1.08824 & -- & -0.7781 & -- & 2.5035 & 19 & 2.8874 & $5.58 \times 10^{-5}$ \\
${\frac{5}{2}}^-$  & -0.30084 & -- & -0.4792 & --  & 3.2565 & 1.5 & 2.5644 & $9.80 \times 10^{-6}$ \\
${\frac{3}{2}}^-$  & 0.4102 & 40 & 0.4226  & 1.2768  & 4.0395 & 225 & 3.2104 & 0.00552  \\
${\frac{3}{2}}^+$  & 0.9412 & 96 & 0.9534 & 129 & 4.3995 & 1530 & 3.9557 &  0.906  \\
${\frac{9}{2}}^-$  & 1.0722 & $< 0.1$ & 2.1528 & $1.08 \times 10^{-7}$ & 4.6195 & - & 5.3930 & $1.26 \times 10^{-9}$ \\
${\frac{3}{2}}^-$  & 1.2356 & 28 & 2.7332 & 0.2923 & 4.8875  & 68 & 5.8526 & $6.78 \times 10^{-5}$  \\
${\frac{7}{2}}^-$  & 1.55366 & 3.4 & 1.2185 & 0.1615 & 5.0715 & 40 & 4.3679 & $1.954 \times 10^{-3}$  \\
(${\frac{5}{2}}^- $)  & 1.5892 & $< 1$ & 3.1504 & 0.1982 & 5.0815$^*$ & $ < 0.6$ & 6.3027 & $6.8 \times 10^{-4}$  \\
${\frac{3}{2}}^+$  & 1.7255 & 6.6 & 4.0680 & 40.226 & 5.2195 & 180 & 7.3661 & 0.0484  \\
${\frac{1}{2}}^-$  & 1.7954 & 32 & 3.5670 & 26.51 & 5.4365 & 30 & 6.6181 & 0.0302  \\ 
${\frac{1}{2}}^+$  & 2.2124 & 124 & 3.0612 & 0.3541 & 5.9595 & 200 & 6.0004 & 0.212  \\ 
(${\frac{5}{2}}^+$)  & 2.7184 & $< 1$ & 2.6958 & $8.3616 \times 10^{-2}$ & 5.9790$^*$& $ < 1.6$ & 5.6928 & $1.25 \times 10^{-4}$  \\
(${\frac{7}{2}}^-$)  & 2.8284 & $< 1$ & 2.4923 & 0.8880 & 6.9455$^*$  & 30 & 5.6906 & $0.0039$  \\
${\frac{5}{2}}^-$  & 3.0221 & 1.38 & 3.7962 & 1.0455 & 6.4265 & 3.8 & 6.5382 & $1.216 \times 10^{-3}$  \\
${\frac{3}{2}}^+$  & 3.0584 & 280 & 4.9729 & 53.98 & 6.7555 & 10 & 7.6222 & 0.0790  \\
${\frac{5}{2}}^+$  & 3.2356 & 0.64 & 3.2894 & 4.8518 & 7.3495$^*$ & 10 & 6.3457 & 0.00808  \\
${\frac{5}{2}}^-$  & 3.2386 & 0.96 & 4.9148 &  0.1685 & 7.7825$^*$ & 11 & 8.0635 & 0.00666  \\
${\frac{3}{2}}^-$  & 3.4154 & 500 & 4.0794 & 0.4399 & 6.1735$^*$ & 4.5 & 7.7045 & $1.46 \times 10^{-3}$  \\
(${\frac{7}{2}}^+$)  & 3.4324 &$< 0.1$ & 3.2902 & $2.58 \times 10^{-3}$ & 6.8705 & 5 & 6.6024 & $1.304 \times 10^{-5}$  \\
${\frac{7}{2}}^-$  & 3.5446 &14.4 & 4.5986 & 1.2018 & 7.4095 & 50 & 7.7624 & 0.01786  \\
${\frac{11}{2}}^-$  & 3.6134 & & 1.4902 & $7.5970 \times 10^{-6}$ & 6.8475$^*$ & $< 5$ & 4.7564 & $5.44 \times 10^{-7}$ \\
${\frac{1}{2}}^+$  & 3.8124 & 90 & 3.7228 & 228  & 7.1495 & 179 & 7.0725 & 267 \\ 
${\frac{1}{2}}^-$  & 3.8469 & 270 & 4.8343 & 21.55  & 7.4745 & - & 7.4661 & 0.0454  \\`
${\frac{3}{2}}^+$  & 3.9264 & 85 & 5.3846 & 0.8878 & 6.8785  & 795 & 8.8502 & 24.7  \\
${\frac{3}{2}}^-$  & 4.0564 & 60 & 5.4326 & 13.8864 & 7.5995$^*$  & 700 & 7.7045 & 8.4175 \\
${\frac{1}{2}}^+$  & 4.1988 & 11.4 & 6.8393 & 99.201 & 7.8155 & 45 & 10.1828 & 366 \\
${\frac{5}{2}}^+$  & 4.2587 & 6.17 & 3.9770 & 25.782 & 7.4695$^*$ & 100 & 7.2445 & 26.53 \\
${\frac{9}{2}}^+$  & 4.3224 & 2.13 & 2.5973 & 0.0711 & 6.8535 & 7 & 5.9439 & 0.3759  \\
\end{tabular}
\end{ruledtabular}
\end{table}
There are clear mismatches in the lists, but it is noteworthy that of the thirty levels listed
for $^{17}$O and $^{17}$F, twenty in $^{17}$O and twenty-four in $^{17}$F, have
matching experimental and MCAS-evaluated partners within one MeV in excitation of each other.
Furthermore, the majority of the larger mismatched pairs lie above 7 MeV in excitation
and we expect that coupling of additional target states to those used would have more 
influence with increasing excitation in the clusters.

A  measure of the over-all agreement is the root-mean-square value, 
\begin{equation}
\label{MSQ}
\mu_N = \sqrt{\frac{\sum_{n=1}^N [E_{exp}(n) - E(n)]^2}{N}} ,
\end{equation}
where $N$ is the number of bound states and resonance centroid energies considered
and $E_{exp}(n)$ and $E(n)$, respectively, are the experimental and calculated
values of the bound and resonance centroid energies in the set.
The root-mean-square value, Eq.(\ref{MSQ}), for the calculated levels in $^{17}$O, 
considering the lowest 30 energy levels is $\mu_{30} = 1.2371$ MeV, 
and with just the lowest 20 levels, $\mu_{20} = 1.1240$ MeV.


To study the mirror system to $n + ^{16}$O, namely $p + ^{16}$O leading to the compound system $^{17}$F,
we use the same parameter set as in 
Table~\ref{param-tab} with the addition of the Coulomb interaction. 
(A Coulomb potential has been generated from the charge distribution assumed for ${}^{16}$O.)
The charge distribution of the protons in $^{16}$O, is described by
 a three-parameter Fermi charge distribution geometry given by 
 \begin{equation} 
 \label{3pF}
 \rho_{ch}(r) = \rho_0\frac{1+w(\frac{r}{R_c})^2}{1+exp(\frac{r-R_c}{a_c})},
\end{equation}
where the parameters $R_c$, $a_c$ and $w$ were obtained from experiment to 
have the values~\cite{Vr87} given in the top of the last column in Table~\ref{param-tab}.
For $^{17}$F, the comparison with experiment shown in Fig.~\ref{Fig3}, is even better than for $^{17}$O, 
giving  $\mu_{30} = 1.0419$ and $\mu_{20} =0.9201$, respectively, 
for the 30 and 20 lowest states. 
However, since a number of the higher-energy levels observed in $^{17}$F have not been given
experimentally known spin-parities, we have made an arbitrary association 
between some measured and calculated levels. 

With respect to the small Coulomb residual displacement energy of 208 keV between 
the experimentally known value for the ground state of $^{17}$F and that calculated by MCAS,
changing to smaller values of $R_c$ and $a_c$, does make the gap smaller. 
But unless quite unrealistic values are used, it is not enough to explain observation.
Possibly a residual gap reflects effects of charge symmetry breaking in the underlying
two-nucleon interactions~\cite{Wu90,Br00}.
This gap is comparable with those found in other mirror systems studied in Ref.~\cite{Br00}.

\subsection{Nucleon scattering cross sections from $^{16}$O.}
\label{subB}

\begin{figure}[ht]
\scalebox{0.7}{\includegraphics*{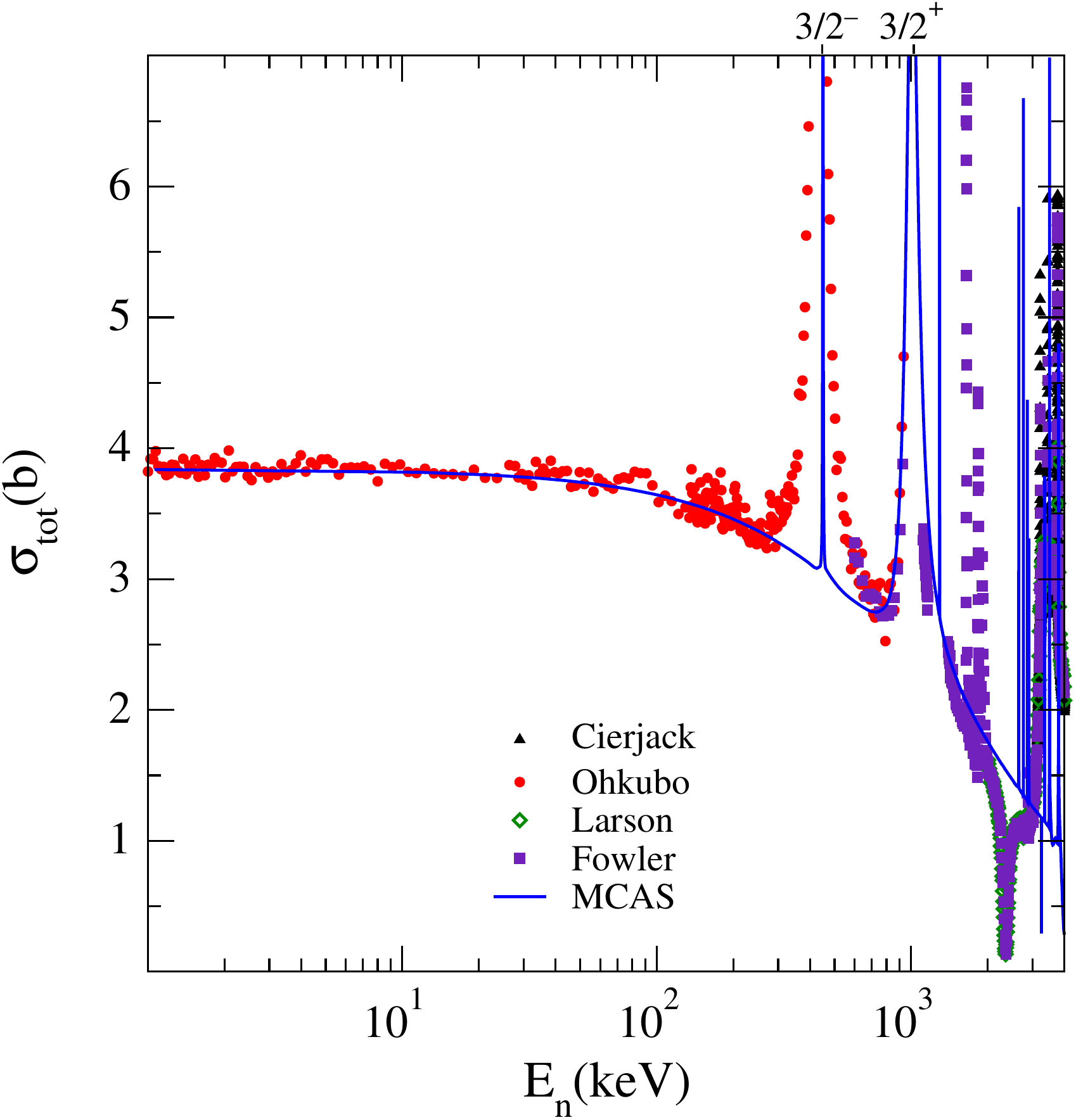}}
\caption{
\label{Fig4} (color online)
Total neutron scattering from $^{16}$O calculated with MCAS (solid line in blue) 
using the parameter set in Table~\ref{param-tab} compared to four data sets.  
The circles are data from  Ohkubo~\cite{Oh87}, the triangles are data from Cierjacks 
{\rm et al}~\cite{Ci80}, the squares are from Fowler {\rm et al}~\cite{Fo73}, and the diamonds 
are from Larson {\rm et al}~\cite{La80}.  The energy scale is $logarithmic$, in units of keV.}
\end{figure}

\begin{figure}[ht]
\scalebox{0.7}{\includegraphics*{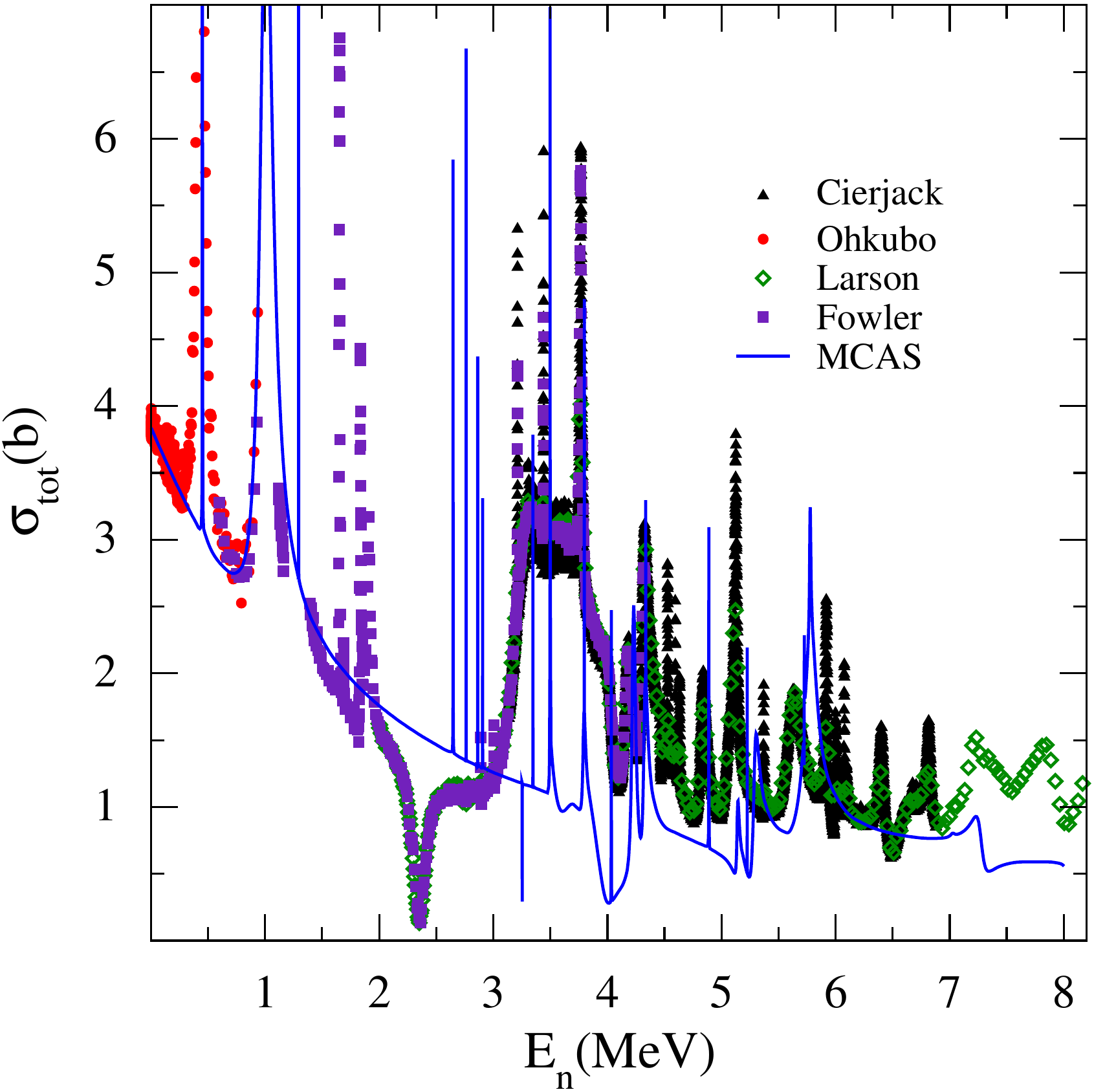}}
\caption{
\label{Fig4-bis} (color online)
Same as in Fig.~\ref{Fig4} but with a linear scale.}
\end{figure}

The total $n$+${}^{16}$O scattering cross section has been calculated using MCAS
as a function of neutron energy to 8.0 MeV using the parameter set in Table~\ref{param-tab}.
In Fig.~\ref{Fig4} the results are compared with data on a logarithmic energy scale.
This emphasizes the very low-energy values and reveals that the calculated cross sections
agree with observation very well at energies $\leq$\nolinebreak 1 MeV. 
The first large resonance, labelled ${\frac{3}{2}}^-$ is in the correct position in the MCAS result,
but its width is much smaller than the experimental one. This resonance decays both 
by $\gamma$- and neutron emission but the radiative width is only $1.8 \pm 0.35$ eV~\cite{Ig92}.  
The neutron width has been assessed~\cite{Ti93} to be $\sim$ 40~keV by analysis of the
elastic neutron scattering cross sections from ${}^{16}$O. 
The higher-energy regime is best shown on a linear scale, as done in Fig.~\ref{Fig4-bis}. 
It shows considerable structure in the MCAS results, and 
resonances are predicted to exist where experiment
reveals some, but the precise matching of resonances in the 3-4 MeV region is not as good as one would like, 
while the backgound cross section is matched fairly well.
As the higher energy region occurs at $\sim$ 8 MeV excitation in the compound 
nucleus, to improve on these cross section results, more target states in ${}^{16}$O are probably 
needed in MCAS calculations.

\begin{figure}[th]
\scalebox{0.7}{\includegraphics*{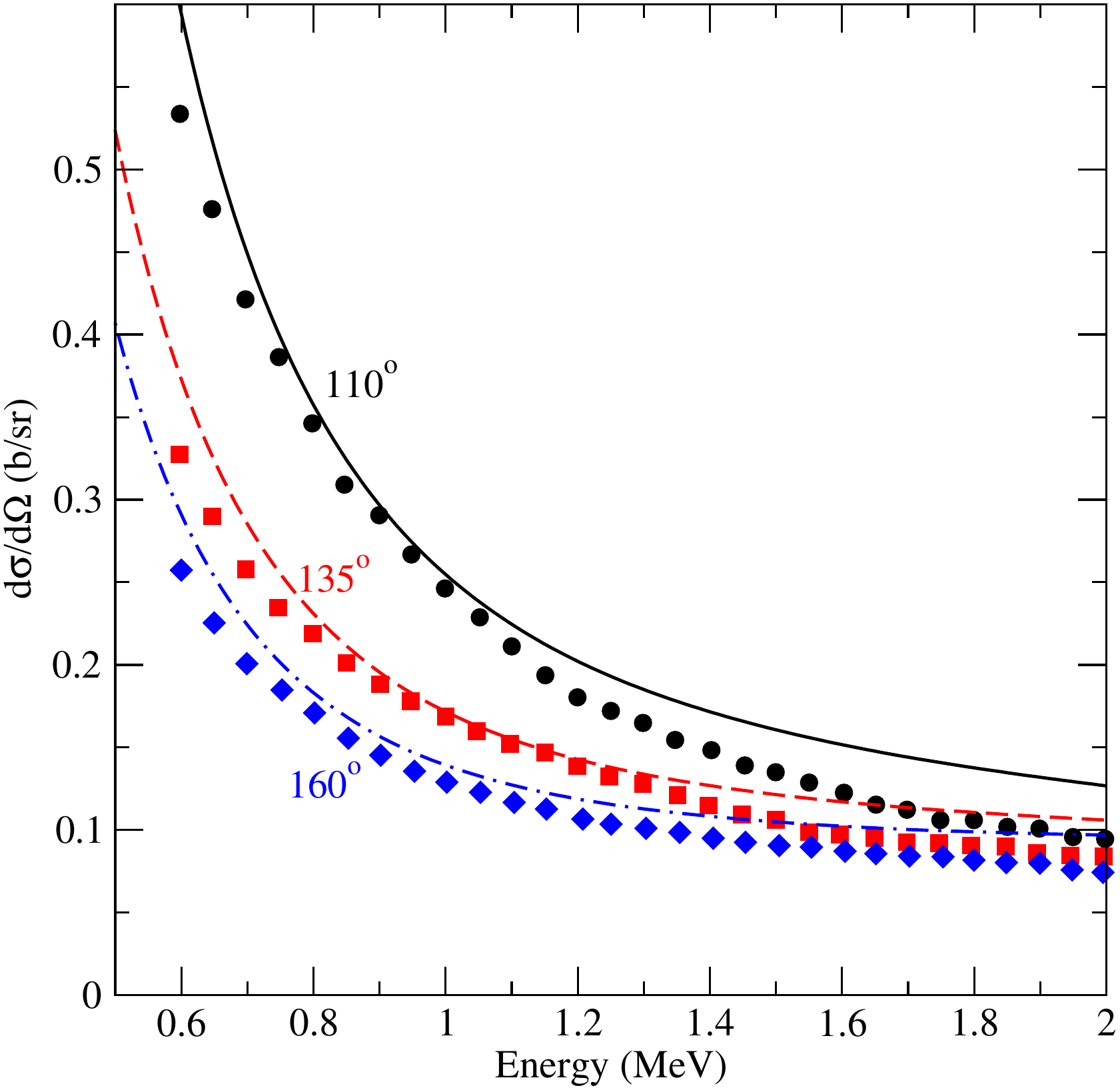}}
\caption{
\label{Braun} (color online)
Differential scattering cross sections of protons from $^{16}$O calculated with MCAS, 
compared to data sets from Braun and Fried~\cite{Br83} 
at three scattering angles. The solid, dashed and dot-dashed lines are the MCAS results at the 
angles shown. The circles, squares and diamonds are data points at the corresponding angles shown.} 
\end{figure}

\begin{figure}[th]
\scalebox{0.7}{\includegraphics*{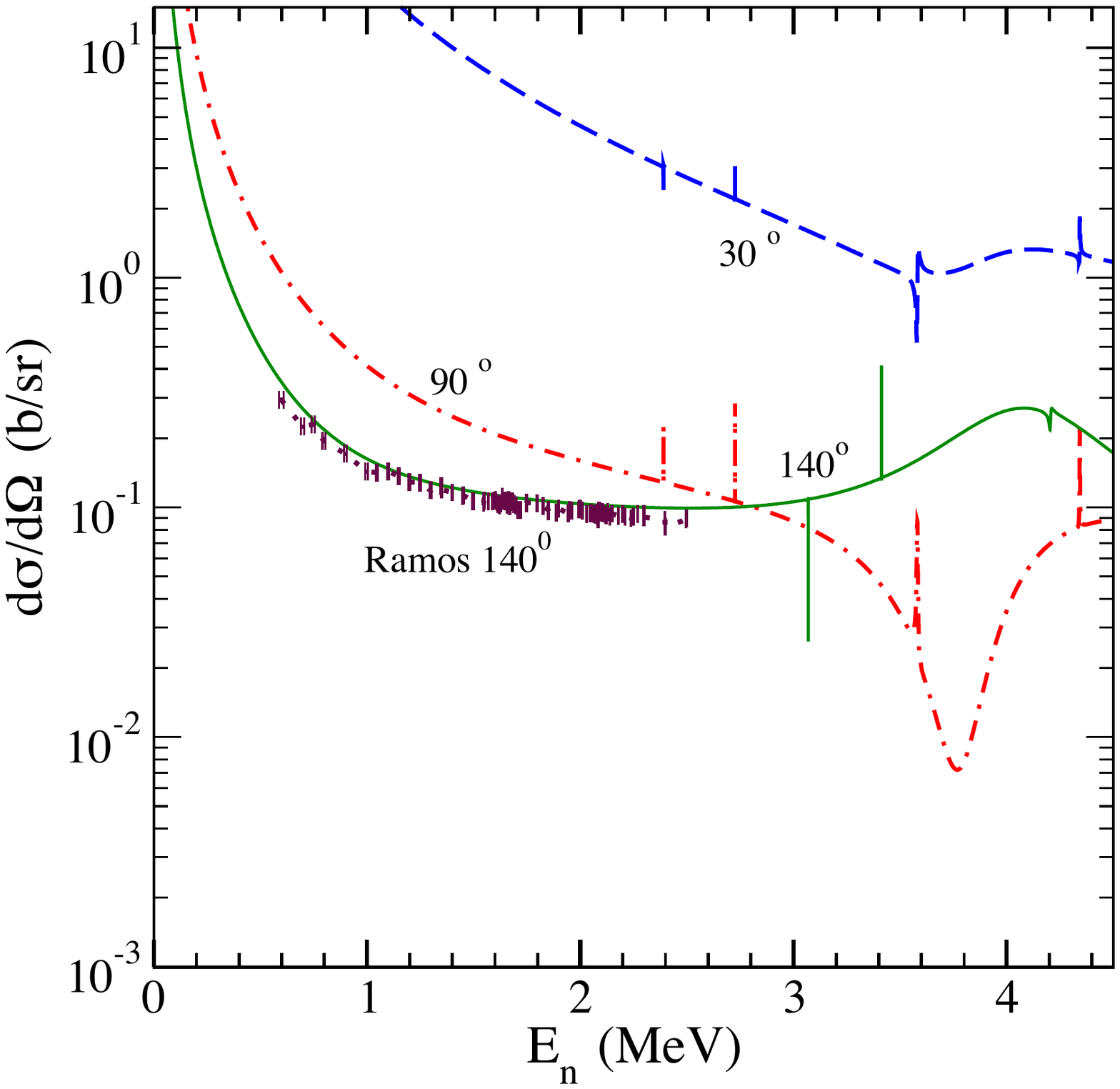}}
\caption{
\label{Ram140} (color online)
Differential scattering cross sections of protons from $^{16}$O calculated with MCAS, 
compared to a data set from Ramos {\it et al.}~\cite{Ra93} 
at scattering angle 140 deg.}
\end{figure}

\begin{figure}[th]
\scalebox{0.7}{\includegraphics*{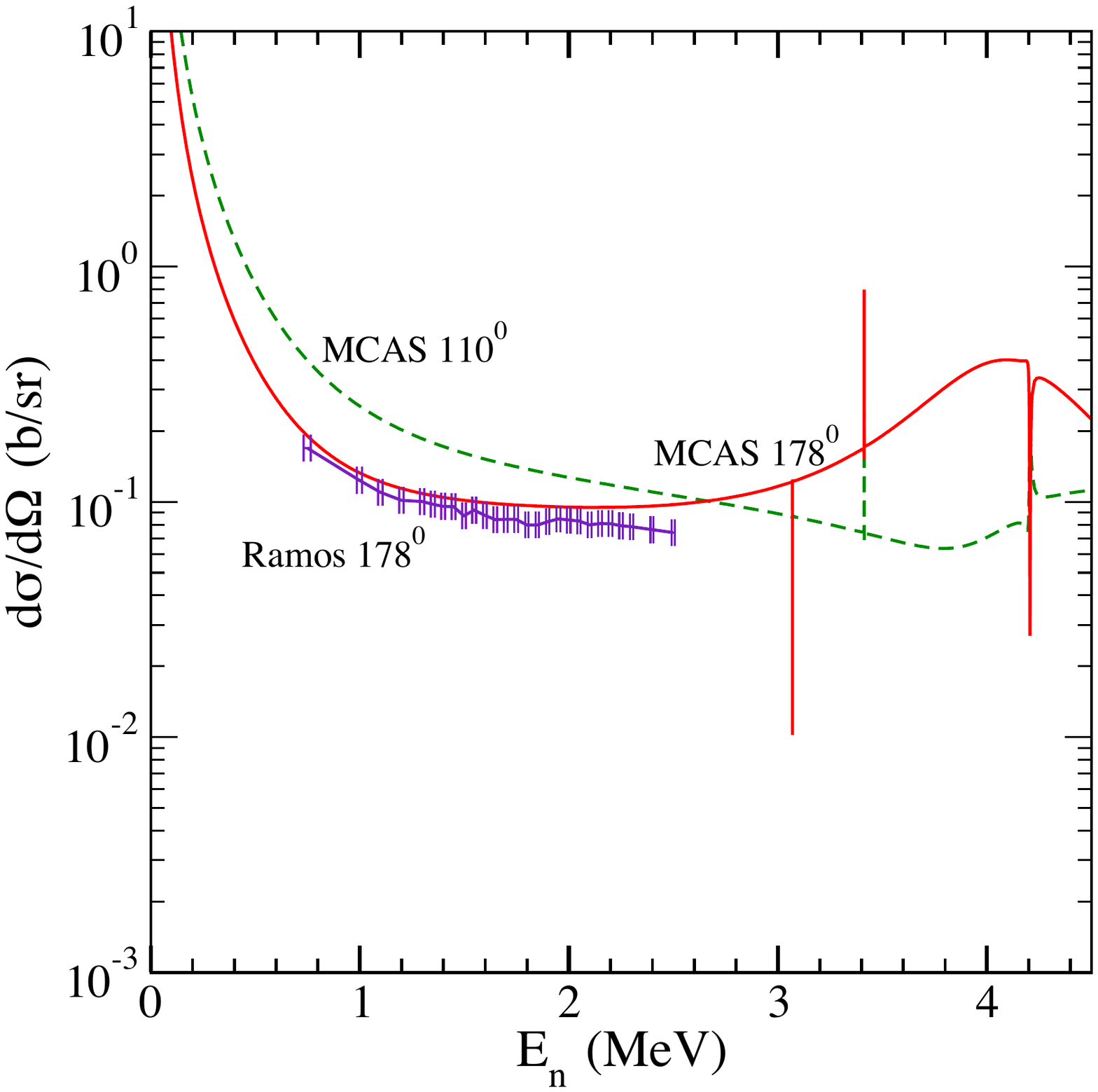}}
\caption{
\label{Ram178} (color online)
As in Fig.~\ref{Ram140}, compared to a data set from Ramos {\it et al.}~\cite{Ra02} 
at scattering angle 178 deg.}
\end{figure}

Next we consider the scattering cross section for $p + ^{16}$O.
In Fig.~\ref{Braun}, differential cross sections at three different scattering angles are
shown as function of energy from 0 to 2.0 MeV with data from Braun and Fried~\cite{Br83}.
The next two Figs., \ref{Ram140} and \ref{Ram178}, show some $p+^{16}$O scattering 
results from Ramos~\cite{Ra93,Ra02}, at two angles, together with MCAS results at the angles 
used in the Ramos work. For comparison, some calculated with MCAS at other angles 
          are shown. There is reasonable agreement between the MCAS results and the
          data, and the calculated results show a measurable variation with energy 
          and angle as well as possible resonance attributes.
Reasonable agreement between MCAS results and data is seen.

\section{Effect of variations of the $\beta_L$ parameters.}
\label{subC}

In Table~\ref{param-tab} we presented the set of parameters that were used for the MCAS calculations of neutron and proton scattering from 
$^{16}$O to obtain the results of Fig.~\ref{Fig3}-\ref{Ram178}, and Table~\ref{Low30-NEW}. Not all the parameters in Table~\ref{param-tab} were treated equally. Those fitted are, essentially, the interaction strengths, ($V_0$, $V_{\ell \ell}$, $V_{\ell s}$, $V_{s s}$), and the  $\beta_2$ and $\beta_3$ coupling parameters.  In contrast, the radius and diffuseness, $R_0$ and $a$, have been held fixed. According to the analysis discussed in Ref.~\cite{Fr15}, $R_0$ denotes the Hamitonian nucleon-nucleus interaction radius, which  is different from the charge radius, $R_c$ of $^{16}$O, taken from ~\cite{Vr87}. In the calculation shown in Table~\ref{Low30-NEW} and Figs.~\ref{Fig3} - \ref{Ram178}, the $\beta_2$ and $\beta_3$  parameters were adjusted to the values given in Table~\ref{param-tab} to get optimal results in the coupled-channel calculations. However, as an alternative to this approach, $\beta_2$ and $\beta_3$ could also be linked to the experimental B(E2) and B(E3) values, which lead to $\beta_2 = 0.362 \pm 0.018$~\cite{SR01,St05,Ish05}, and similarly to a value of 0.6 for the octupole coupling $\beta_3$~\cite{SPEA89}. Therefore, with the aim to consider this alternative option, we made new calculation fixing $\beta_2 = 0.36$ and 
$\beta_3 = 0.6$ and refitting the remaining adjustable parameters. The varied list of parameters values is reported in Table~\ref{Newtab3}.

\begin{table}[h]
\begin{ruledtabular}
\caption{\label{Newtab3}
New parameter values for $n$+${}^{16}$O MCAS cluster structure. }
\begin{tabular}{cccccc}
$V_x$ (MeV) & $P = -$ & $P = +$ &  Geometry & value &\\
\hline
$V_0$ & $-$45.0 & $-$45.0 & $R_0$ & 3.15 fm &\\
$V_{ll}$ &\ 0.55 &\ $-$0.216 & $a$ &  0.65 fm &\\
$V_{ls}$ &\ 8.71 &\ 8.71 & $\beta_2$ &  0.36 & \\
$V_{ss}$ &\ 2.0\ &\ 1.9\  & $\beta_3$ & 0.6 &\\
\hline
$I_n^\pi$ & $E_n$ (MeV) & $0s_{\frac{1}{2}}$ & $0p_{\frac{3}{2}}$ &
$0p_{\frac{1}{2}}$ & $0d_{\frac{5}{2}}$ \\
\hline
$0^+_1$ &  0.0 & 10$^6$ & 10$^6$ & 10$^6$ & 0.0\\
$0^+_2$ & 6.049 & 10$^6$ & 10$^6$ & 0.0 & 0.0 \\
$3^-_1$ & 6.13  & 10$^6$ & 10$^6$ & 5.0  & 1.0\\
$2^+_1$ & 6.92  & 10$^6$ & 10$^6$ & 0.5 & 0.0 \\
$1^-_1$ & 7.12 & 10$^6$ & 10$^6$ &  5.0 & 1.0\\
\end{tabular}
\end{ruledtabular}
\end{table}

\begin{figure}[ht]
\scalebox{0.75}{\includegraphics*{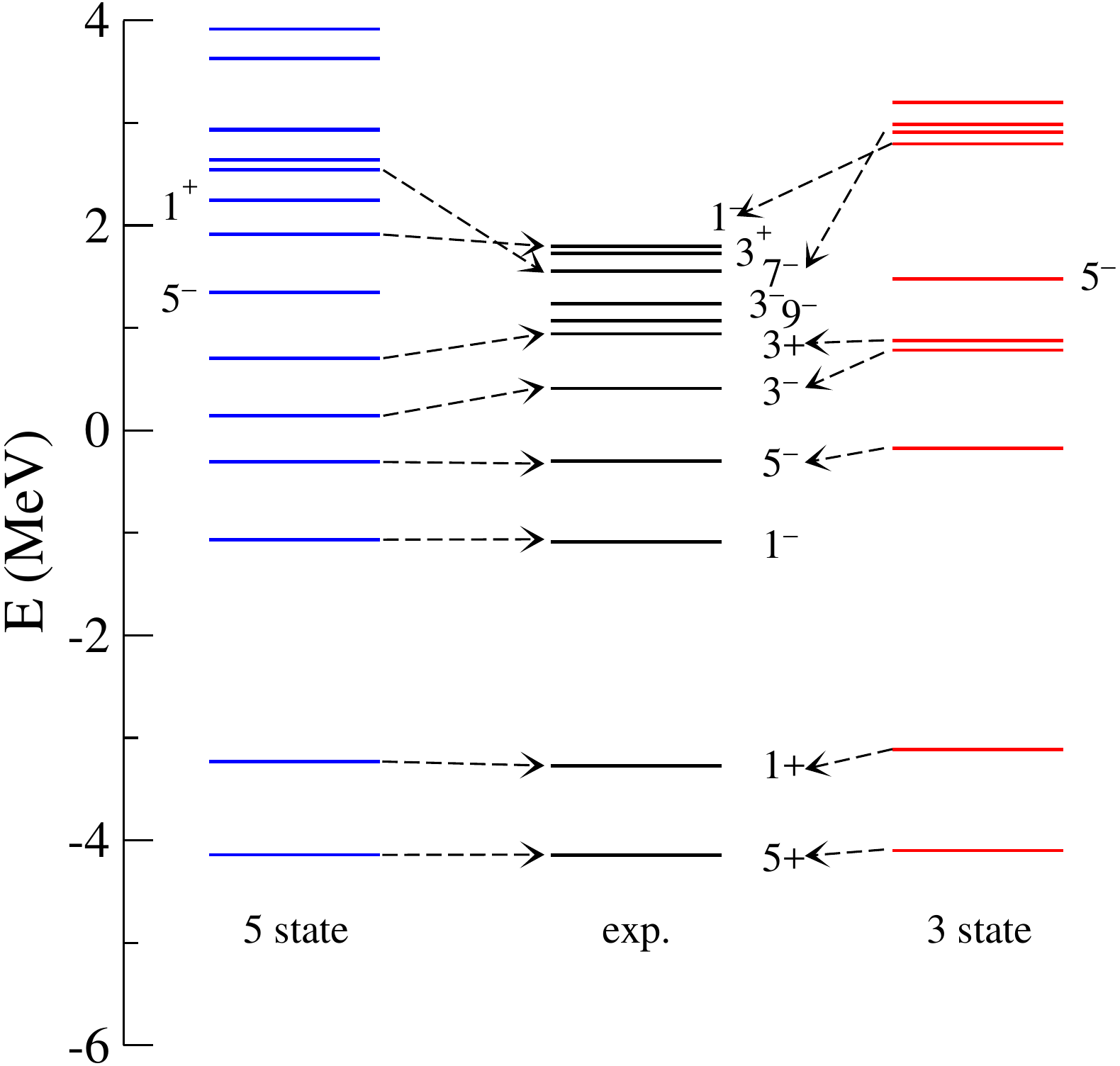}}
\caption{\label{Fig6} (color online)
Spectra of $^{17}$O, left and right are calculated with MCAS including/excluding the two states $0^+$ and $1^-$.
The middle column represents  the  experimental spectrum. The calculations have been performed with $\beta_2=0.36$ and $\beta_3=0.6$, according to Table~\ref{Newtab3}.
The numbers by the levels are twice the spin of the level, and the superscript  indicates the parity.  The zero on the scale is at the 
$n + ^{16}$O threshold.}
\end{figure}
\begin{figure}[ht]
\scalebox{0.75}{\includegraphics*{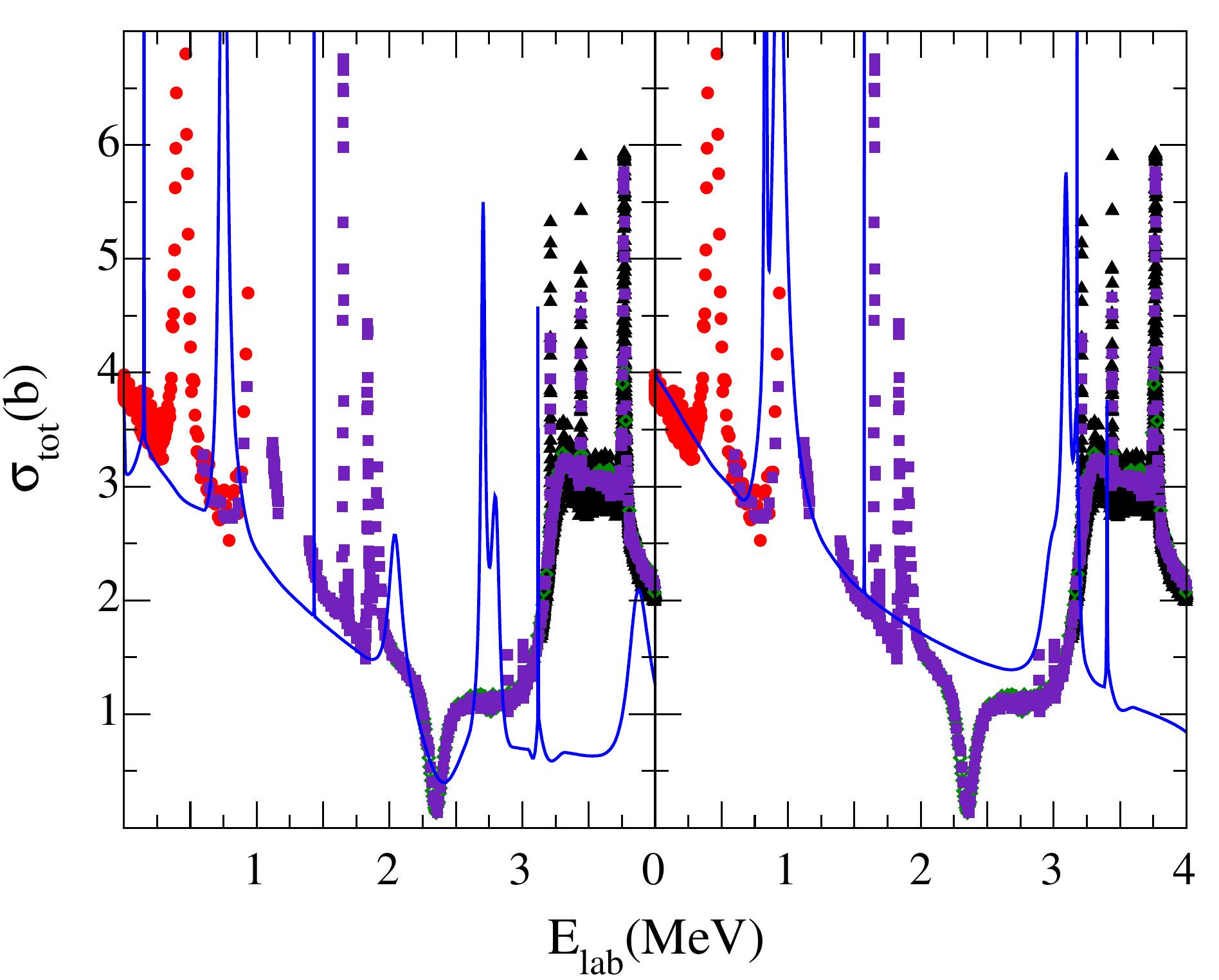}}
\caption{\label{Fig7} 
 (color online)
Total neutron scattering from $^{16}$O calculated with MCAS (solid line), utilizing the
parameter set in Table III (using $\beta_2=0.36$ and $\beta_3=0.6$). Left panel refers to a five-state calculation while right 
panel to a three-state calculation ($0^+$gs, $3^-$, $2^+$). 
The experimental data are the same of Fig.~\ref{Fig4}}. 

\end{figure}

A distinctive feature of the model couplings discussed in Appendix B is that the $^{16}$O $0^+$ and $1^-$ excited states
are coupled through second-order couplings of quadrupole-octupole vibrations. This choice is very specific for the schematic model  considered herein. Other possible excitation/de-excitation modes (e.g., monopole or dipole couplings) are not contemplated in the model given in Appendix B. To estimate the effect of those two states and their couplings, we compare the full (five state) calculation with a calculation where the couplings to the $0^+$ and $1^-$ excitations have been removed. This alternative calculation is denoted as a 3-state calculation ($0^+$gs, $3^-$, $2^+$) in Figs.~\ref{Fig6} and ~\ref{Fig7}.  

In Fig.~\ref{Fig6} we illustrate the bound and resonant spectra of ${}^{17}$O when the $\beta$ couplings have been
set at the adopted values. The results on the left column 
refer to the five-state calculation, while the results on the right column 
refer to the corresponding three-state calculation. The middle column 
contains the known experimental data. Note that the two calculated results are quite similar except that 
with the three-state calculation we completely miss the second excited 
state $(\frac{1}{2})^-$.  
In Fig.~\ref{Fig7}, we show the neutron-oxygen total scattering cross section obtained from the same variation of the $\beta_2$ as used for Fig.~\ref{Fig6} for the two model (three-state and five-state) calculations. There are small but significant differences between the calculations shown in Fig.~\ref{Fig4}  and in Fig.~\ref{Fig7}; the former showing a near-perfect agreement between data and MCAS 
(with  $\beta_2 = 0.21$ and $\beta_3 = 0.42$). It is not clearly understood if one should use for these $\beta_L$ values those deduced from (electromagnetic) experiments with $^{16}$O, as in Fig.~\ref{Fig7}, or if the nucleon interaction leads to some modification of these deformation values in the coupled-channel dynamics. After all, the interaction radius itself is affected by the presence of the incoming nucleon, as discussed in Ref.~\cite{Fr15}, and leads us to use a value which is different from the charge radius of the $^{16}$O target. The same could happen for the $\beta$ parameters.

\section{Summary of EM transitions}
\label{subEM}

Finally, we present here a summary of EM transitions in $^{16}$O that can be obtained with the collective model we employ.
In particular, the $E0$, $E2$, and $E3$ transition properties between some of these states
have been assessed; data on those of prime interest are as listed in
Table~\ref{TabEM1}.
\begin{table}[h]
\begin{ruledtabular}
\caption{\label{TabEM1}
Electromagnetic transition properties in ${}^{16}$O}
\begin{tabular}{ccccc}
Type & Transition & model result & exp. value & Reference\\
\hline
$\rho^2(E0)$ & $0^+_2 \to 0^+_1$ & 0.026 & 0.153 & \cite{Ki05}\\
$B(E2)$ & $0^+_1 \to 2^+_1$ & 40.6 (e$^2$-fm$^4$) & 23-51 (e$^2$ fm$^4$) & 
\cite{Ra01}\\
$B(E3)$ & $0^+_1 \to 3^-_1$ & 900 (e$^2$-fm$^6$) & 400-1550 (e$^2$-fm$^6$)& \cite{SPEA89}\\
\end{tabular}
\end{ruledtabular}
\end{table}

The $E2: 0^+_1 \to 2^+_1$ values given in~\cite{Ra01} span
a wide range and all have been extracted from experimental
data.  However, the value 40.6 e$^2$-fm$^4$ has been adopted.
With that value, and assuming for ${}^{16}$O a 
uniform spherical charge density in its ground state
(radius $R = 1.2$A$^{1/3}$ fm.), the base vibration model gives  the
deformation parameter as~\cite{Ra01}
\begin{equation}
\beta_2 = \frac{4\pi}{3Z R_0^2} \sqrt{B(E2: 0_1^+ \to 2_1^+)} 
= 0.36 \ {\rm for\ } {}^{16}{\rm O} .
\end{equation}

With the same model geometry, for ${}^{16}$O, and using 
the link~\cite{Da68}  between $\rho(E0: 0_2^+ \to 0^+_1)$ 
and the $B(E2: 2_1^+ \to 0_1^+)$,
\begin{equation}
\rho(E0: 0_2^+ \to 0^+_1)\ =\ 
\sqrt{10}\ \frac{4\pi}{3} \frac{1}{Ze^2 R_0^4}
B(E2: 2_1^+ \to 0_1^+) , 
\end{equation}
and so the square is
\begin{equation}
\rho^2(E0: 0_2^+ \to 0^+_1) \ =\
 \frac{320\pi^2}{9 Z^2 e^4 R_0^8}
\left[B(E2: 2_1^+ \to 0^+_1)\right]^2
\ =\ 0.026:
\label{mylink}
\end{equation}
a factor of $\sim 6$ smaller than observed. 
Of course the structure model considered is simplistic, with phenomena
like shape coexistence and non-collective attributes known to influence 
monopole strengths.  For the same reason, our simple vibration model gives zero 
for the direct isoscalar E1 matrix element.

Likewise the $E3: 0^+_1 \to 3^-_1$ values given in~\cite{SPEA89} span
a wide range and all have been extracted from experimental
data.  We use an average value of 900 e$^2$-fm$^6$ with 
which the basic vibration model for ${}^{16}$O gives  the
deformation parameter, $\beta_3 = 0.6$. 

With reference to $^{16}$O, it is well known that a $E1$ transition from a 1− to the 0+ gs state has been observed with an extremely small transition probability. (See Refs.~\cite{Swa70,Mis75,Amo88} and references therein.) However, this is not accounted for in this study, or in most other investigations to date. (See Ref.~\cite{Ari75} for an investigation into the underlying causes.)
%

\section{conclusions}
\label{sec6}

The MCAS method for nucleon-nucleus scattering studies was developed and first used for neutron scattering from the well-known nucleus $^{12}$C~\cite{Am03}. 
The structure of that target nucleus was described by a rotational model with a deformed Fermi function, with the deformation specified by a 
$\beta_L$ value. The parameters of the system were chosen to obtain a very good description of the neutron-$^{12}$C elastic scattering cross section. This study also yielded a good description of the energy levels in $^{13}$C, both bound states and resonances. Since that first result, a number of other nucleon + nucleus systems have been evaluated and studied by the MCAS method, all using the rotational model.

In this work the MCAS method has been applied for the first time with a vibrational model for the target nucleus, 
to study nucleon scattering on the $ ^{16}$O nucleus. The spectra of $^{17}$O and of $^{17}$F have been evaluated using the MCAS approach, 
treating these nuclei as $n + ^{16}$O, and $p + ^{16}$O compounds, respectively. 

As the main result, we have shown that, with this approach, it is possible to describe the very low-energy cross section for neutron and proton elastic scattering on $^{16}$O with a coupled-channel model that takes into account the excitation dynamics of the low-lying collective 
states of $^{16}$O. The calculation performed and the results obtained show that the approach has potential interest 
for any application where the determination of low-energy cross section are of great importance. For instance, the low energy regime is of import for capture cross sections~\cite{Ro88},  as well as for nuclear reactor physics applications~\cite{Ro14}.

It must be observed that the vibrational coupled-channel model in its present form is still at a preliminary stage, and that a variety 
of improvements can be performed in future studies. For example, the use of fit parameters typical of a macroscopic theory 
can be reduced if not fully removed if we use insights coming from microscopic theories. This is especially so if one uses ground state densities coming from folding model calculations, and in a similar manner one derives the transition densities for the coupling interactions.
However, at present there is no microscopic (or microscopically inspired) theory that works so well in this low-energy scattering regime.
Another improvement could consider couplings to the excited $0^+$ and $1^-$ states derived directly from first order transition of monopole, dipole structure, while in the present model we take into account for these states only second order transitions of quadrupole plus octupole type.
With these caveats, we have described the neutron or proton $+^{16}$O coupled-channel dynamics using the five lowest excited states in $^{16}$O with the interaction potentials specified by a collective vibration model for the target states. With those interactions, the Pauli principle was satisfied by using the orthogonalizing pseudo-potential scheme~\cite{Ca06}, and then,  good agreement between theory and data at low energies was found.  While there remain discrepancies, such as a small residual displacement energy, of thirty levels listed 
in Table~\ref{Low30-NEW} for $^{17}$O and $^{17}$F, twenty in $^{17}$O and twenty-four in $^{17}$F have matching MCAS evaluated partners within one MeV in excitation of each other.

The total elastic scattering cross section for neutrons on $^{16}$O is a near perfect match to data up to 1 MeV of excitation, except for the widths of the first two peaks around 1 MeV. At higher energy there are additional resonances in the MCAS results, which, however, only approximately match available data. For the scattering of low-energy protons from $^{16}$O, differential cross sections only exist at fixed scattering angles. Our calculated results agree very well with measured ones.

\appendix

\section{Shell model considerations for ${}^{16,17}$O and ${}^{17}$F.}
\label{Appx1}

If ${}^{16}$O is considered to be a doubly-magic nucleus, 
in its ground state the $0s_{\frac{1}{2}}$ and both orbits in the $0p$-shell 
will be fully occupied and that state predominantly would be spherical in shape.
The two nuclei,  ${}^{17}$O and ${}^{17}$F, often have been considered as a single 
nucleon outside an $^{16}$O core, 
and as mirror nuclei with the first three positive-parity states reflecting 
the single particle energies of the $0d_{\frac{5}{2}}$, $1s_{\frac{1}{2}}$, 
and $0d_{\frac{3}{2}}$ levels in the $(0d1s)$-\nolinebreak shell model. 
But the model for each nucleus is not 
so simple: in a $(0+2)\hbar\omega$ shell-model prescription there is significant 
admixing of $2\hbar\omega$ components, $\sim 25\%$, in the ground states. 
This largely stems from 2p-2h components giving rise to additional nucleons in the 
$(0d1s)$-shell. With this in mind, it is instructive to compare the extreme 
shell-model picture, with one particle in the $(0d1s)$-shell, or the more general 
$(0+2)\hbar\omega$ model, to the collective model description contained in 
the MCAS theory~\cite{Am03}, which describes low-energy nucleon-nucleus scattering, 
and the spectrum of the compound system (both bound-states and resonances). 
However, it is well known that the description of the spectrum of $^{16}$O requires 
a  $4\hbar\omega$ shell model at the minimum \cite{Br66,Ha90,Ka96}.
We discuss aspects of both the $2\hbar\omega$ and $4\hbar\omega$ shell-model
results for $^{16}$O to frame discussion of the $2\hbar\omega$ results we have 
been able to obtain, so far, for the mass-17 systems.

Haxton and Johnson~\cite{Ha90} made a $(0+2+4)\hbar \omega$ shell model calculation
of the spectrum of ${}^{16}$O.  They used  a two-nucleon  interaction that consisted of
\begin{itemize}
\item The Cohen and Kurath (8-16)2BME~~\cite{Co65} for the $0p$-shell
\item The Brown and Wildenthal interaction~\cite{Br88} for the $(0d,1s)$-shell 
\item The Millener-Kurath interaction~\cite{Mi75} for the $(0p,0d,1s)$ cross shell-elements, and
\item The bare Kuo $g$-matrix for the $2\hbar \omega$ interaction \cite{Ku66, Ku68}.
\end{itemize}
Every other matrix element necessary to specify the interaction in the complete model space was set to zero. Further, those matrix elements which gave rise to the violation of the Hartree-Fock condition were also removed. In that sense, the interaction was not complete for the model space assumed. Nevertheless, the spectrum they obtained was reasonable and confirmed the Brown and Green result. An extension of that shell model calculation to include negative parity states~\cite{Ka96} also found reasonable agreement for the states in the spectrum. While the single particle basis assumed was complete for the $(0+2+4)\hbar\omega$ space for the calculation of the positive parity states, there was one restriction in the calculation of the negative parity states, which was done in the same single particle basis. That restriction did not allow for the single-particle excitations to the $0i1g2d3s$ shell.

\begin{figure}[h]
\scalebox{0.65}{\includegraphics*{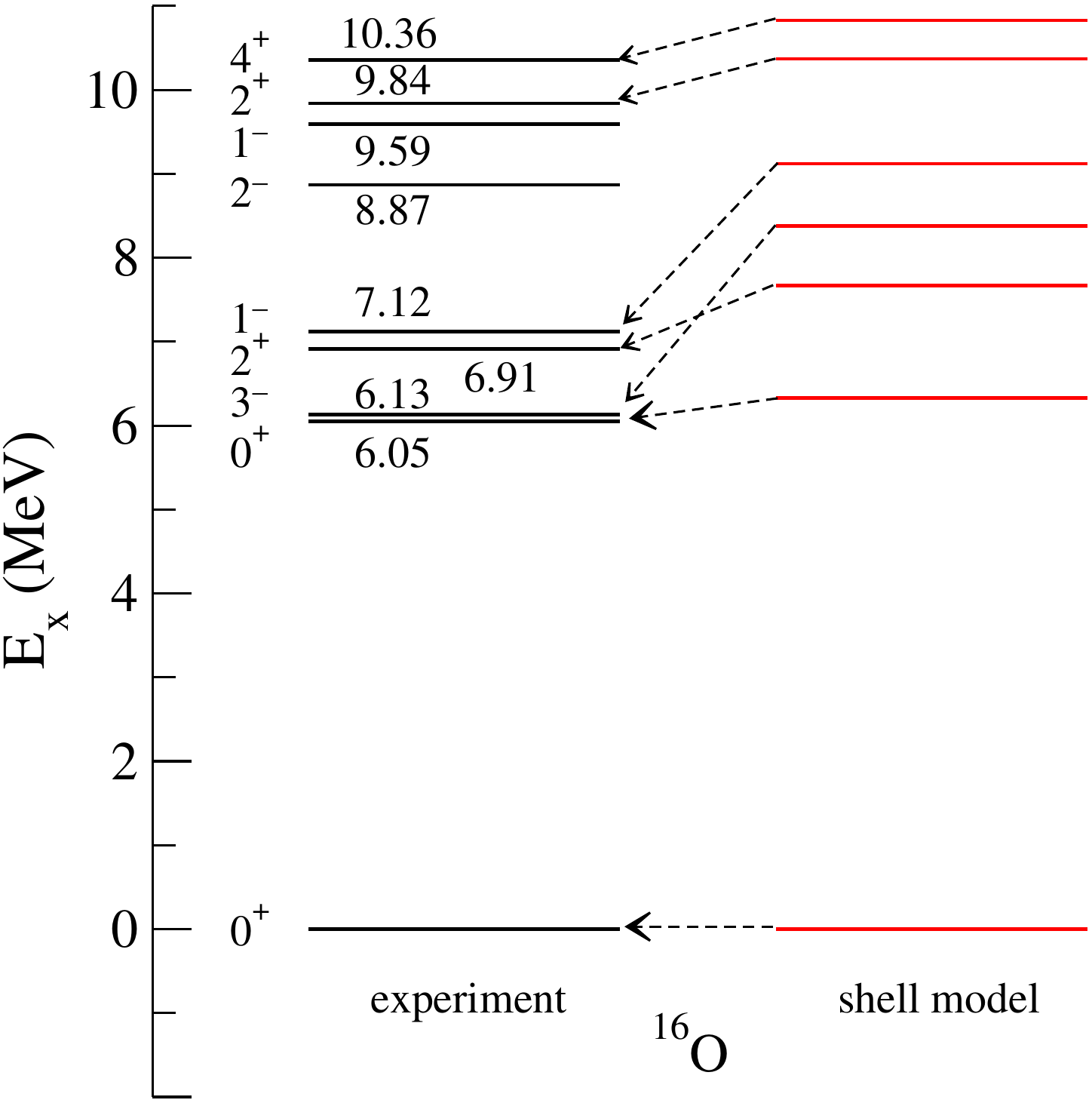}}
\caption{\label{Fig1}(color online)
The low-lying spectrum of ${}^{16}$O. 
The experimental energies~\cite{Ti93} are compared with the shell-model results 
found using the Haxton and Johnson interaction.}
\end{figure}

Haxton and Johnson sought to determine whether the Brown and Green model~\cite{Br66},
which placed importance on inclusion of $4\hbar \omega$ components in the wave functions for
${}^{16}$O, could be reproduced with a microscopic shell-model calculation.
With this scheme, the states of ${}^{16}$O have been determined using the
Haxton version of the GLASGOW shell model program~\cite{Ka96} and 
the results of that~\cite{Ka96} are shown in Fig.~\ref{Fig1}.
Therein, while the positive parity states were evaluated in a complete 
$(0+2+4)\hbar \omega$ space using the Haxton and Johnson interaction, the negative 
parity states were evaluated in a (restricted) $(1+3+5) \hbar \omega$ space, using 
the same interaction. Both calculations used the single particle basis from the 
$0s$-shell up to, and including, the $(0h1f2p)$-shell.
The restriction placed in the $(1+3+5) \hbar \omega$ space is that single-particle 
excitations from the $0p$-shell up to the $(0i1g2d3s)$-shell were excluded. 
This restriction does not guarantee complete removal of center of mass spuriosity but, 
as the center of mass energy for all states obtained in the model is 19.19 MeV, 
there is very little  spuriosity in the specified low-lying states.

All positive parity states displayed in Fig.~\ref{Fig1} are well reproduced by
the calculation, but the $3^-$ and $1^-$ are not, lying about 2 MeV above the experimental  values. 
The predicted energies of the $2^-$ and $1^-_2$ states are at 12.67 and 15.97 MeV, respectively, 
and so are not shown in the figure. 

A complete $(0+2)\hbar \omega$ calculation, using the MK3W interaction, was made for the positive
parity states of ${}^{16}$O as well. That calculation placed all excited states above 20 MeV,
indicating the importance of including $4\hbar\omega$ components to give a sensible mixing of 
$2\hbar\omega$ and 4$\hbar\omega$ components when a $2\hbar\omega$ interaction is involved; 
bringing the energies of states into better agreement with experiment.
The summed shell occupancies (proton and neutron are identical) of the ground state in ${}^{16}$O
from the two shell model calculations are listed in Table~\ref{O16-occc}.

\begin{table}[h]
\begin{ruledtabular}
\caption{\label{O16-occc}
Shell occupancies (proton+neutron) in the ground state of ${}^{16}$O.}
\begin{tabular}{ccccccccccc}
\hline 
Orbital & $0s_{\frac{1}{2}}$ & $0p_{\frac{3}{2}}$ & $0p_{\frac{1}{2}}$ & $0d_{\frac{5}{2}}$ & $0d_{\frac{3}{2}}$ & $1s_{\frac{1}{2}}$ & $0f_{\frac{7}{2}}$ & $0f_{\frac{5}{2}}$ & $1p_{\frac{3}{2}}$ & $1p_{\frac{1}{2}}$ \\
$(0+2)\hbar \omega$ & 3.999 & 7.741 & 3.788 & 0.283 & 0.135 & 0.021 & 0.0 & 0.0 & 0.028 & 0.005\\
$(0+2+4)\hbar \omega$ & 3.996 & 7.319 & 3.262 & 0.831 & 0.441 & 0.138 & 10$^{-3}$ & 7x10$^{-4}$ & 0.003 & 0.002
\end{tabular}
\end{ruledtabular}
\end{table}

Higher orbits in the $(0+2+4)\hbar\omega$ space have occupancies less than $10^{-5}$ nucleons.
From these numbers it is clear that the significant populations in the $(0d1s)$-shell
and the lack of population 
in the higher shells indicates that the ground state is essentially of $(0+2)\hbar\omega$
character, though the distribution in the lower shells is affected by the $4\hbar\omega$
contributions.
\begin{figure}[th]
\scalebox{0.65}{\includegraphics*{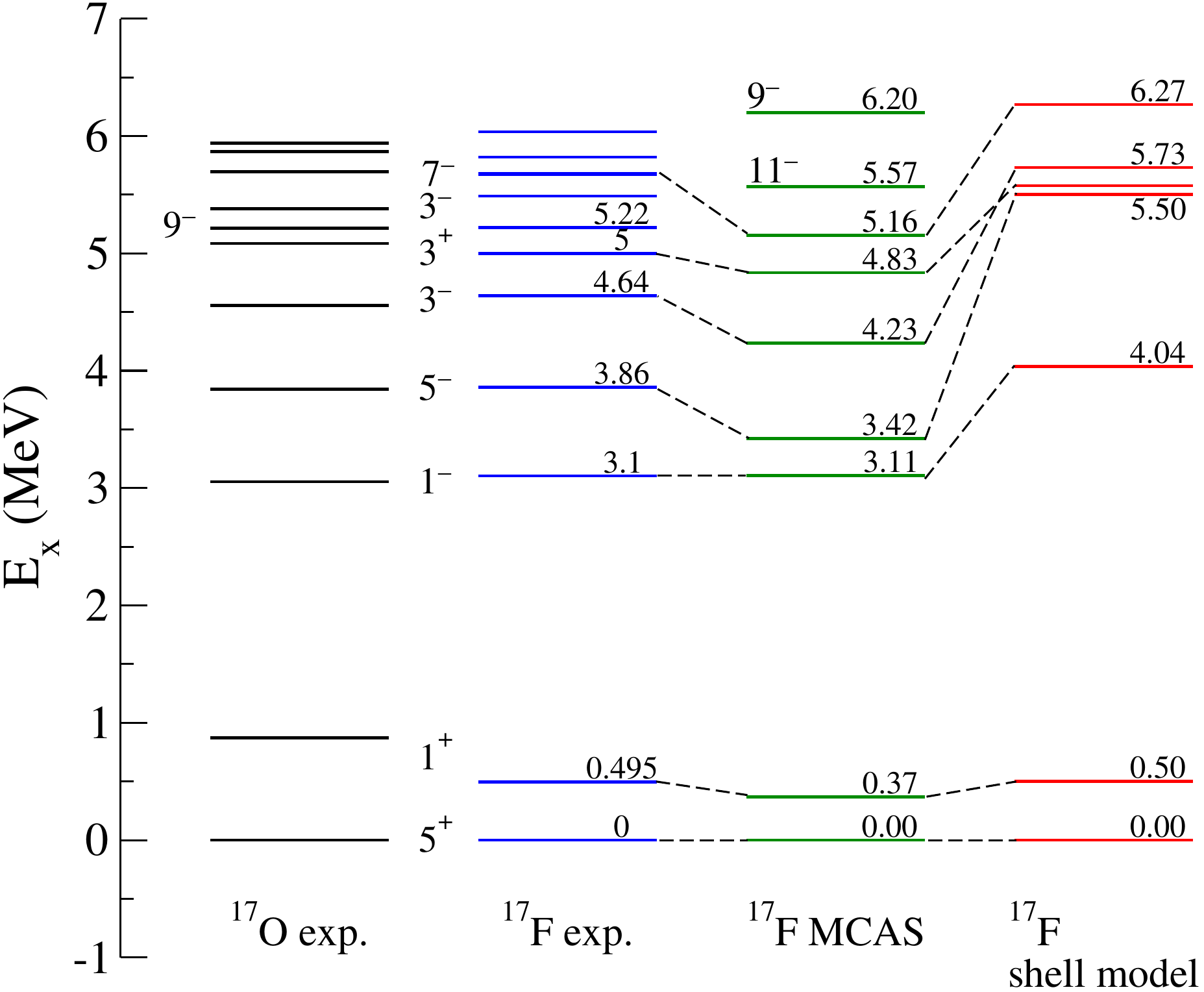}}
\caption{\label{Fig2} (color online) Spectra for $^{17}$O and $^{17}$F \cite{Ti93}, 
with zero energy corresponding to the ground states of each. 
The state labels denote $2J^{\pi}$.}
\end{figure}
As the ground state of ${}^{16}$O is so dominantly of $(0+2)\hbar\omega$ character,
we have calculated the spectra of $^{17}$O and $^{17}$F in that model space,
for the positive parity states, and in a restricted $(1+3)\hbar\omega$  space for the negative 
parity states. In both sets of calculations all shells from the $0s$ to the $(0f1p)$ are used, 
with all particles active.  In these cases the spectra were found again using the OXBASH 
program but with the WBP interaction of Warburton and Brown \cite{Wa92}. 
The resultant shell model spectrum, together with the known spectra for $^{17}$O and
$^{17}$F \cite{Ti93}, is shown in Fig.~\ref{Fig2}.
It is clear that the spectrum obtained from the shell model compares reasonably well with both spectra. Discrepancies between the model and the known spectra may be due to limitations in the model space and/or the underlying limitation on the ground state of $^{16}$O. 
Nevertheless, this result serves to illustrate that the extreme single-particle picture of the mass-17 system is too simplistic. 
It points to the need for a coupled-channel description of the nuclei, with a possibly extended set of (${^{16}}$O) target states to be included in the coupling scheme.

Fig.~\ref{Fig2} compares the low--energy shell model spectrum of $^{17}$F with that of MCAS, using parameters as per Table~\ref{param-tab}, except with $V_0^- = -47.89$ MeV and $V_0^+ = -50.062$ MeV. This small change is made to account for the slight overbinding observed upon use of mirror symmetry.

\section{The vibration model for coupled-channel potentials}
\label{Appx2}

The surface of a liquid drop of incompressible fluid that can be slightly 
deformed is represented as
\begin{equation}
R(\theta \phi)  = R_0 \left[ 1 +
\sum_{\lambda > 1, \mu} \alpha_{\lambda \mu}^\star
Y_{\lambda \mu}(\theta \phi) \right] \ = R_0 \left[1 + \varepsilon \right].
\label{dropsur}
\end{equation}
With this specification of the nuclear surface, expansion to second 
order in the coupling of a function gives,
\begin{equation}
f(r) = f_0(r) + \varepsilon \left(\frac{\partial f(r)}{\partial \varepsilon} \right)_0 
+ \frac{1}{2} \varepsilon^2 \left(\frac{\partial^2f(r)}{\partial \varepsilon^2} \right)_0\ . 
\end{equation}
Then, with $\varepsilon$ as identified by Eq.~(\ref{dropsur}), and treating $R(\theta, \phi)$ 
as the variable in \linebreak $f(r) = f(r-R(\theta, \phi))$,
\begin{eqnarray}
f(r) = f_0(r) &-& R_0 \sum_{\lambda \mu} \alpha_{\lambda \mu}^\star
Y_{\lambda \mu}(\theta, \phi) \left(\frac{\partial f(r)}{\partial r}\right)_0\nonumber\\ 
&+& \frac{1}{2} 
R_0^2 \sum_{l_1 m_1 l_2 m_2} \alpha_{l_1 m_1}^\star \alpha_{l_2 m_2}^\star
Y_{l_1 m_1}(\theta, \phi) Y_{l_2 m_2}(\theta, \phi) \left(\frac{\partial^2f(r)}{\partial r^2} 
\right)_0\ .
\label{Vib-fst}
\end{eqnarray}
Similar forms exist for $g(r) = \frac{1}{r}\frac{\partial f(r)}{\partial r}$, the usual function
taken for spin-orbit terms. 
Therein, and in all that follows, it is presumed that summation of the
expansion labels of the generalised coordinates, and subsequently of the
angular momentum quantum numbers of the phonon creation/anihilation operators 
derived from them, exclude dipole forms to ensure that there is no spurious centre-of-mass
motion associated with a scalar interaction.

The product of two generalised coordinates that satisfy the spherical harmonic 
condition can then  be written as,
\begin{align}
\alpha_{l_1 m_1}^\star \alpha_{l_2 m_2}^\star
=&\ \sum_{\nu_1 \nu_2}  \delta_{m_1 \nu_1} \delta_{m_2 \nu_2}
\alpha_{l_1 \nu_1}^\star \alpha_{l_2 \nu_2}^\star
= \sum_{\lambda \mu}
\left< l_1 l_2 m_1 m_2 \vert \lambda \mu \right>
\left[\alpha_{l_1}^\star \otimes \alpha_{l_2}^\star \right]_{\lambda
\mu}
\nonumber\\
&\hspace*{0.5cm} 
\left[\alpha_{l_1}^\star \otimes \alpha_{l_2}^\star \right]_{\lambda
\mu}\
=\ \sum_{\nu_1 \nu_2} 
\left< l_1 l_2 \nu_1 \nu_2 \vert \lambda \mu\right>
\ \alpha_{l_1 \nu_1}^\star \alpha_{l_2 \nu_2}^\star .
\label{contract-aa}
\end{align}
This form is convenient since 
$\left[\alpha_{l_1}^\star \otimes \alpha_{l_2}^\star \right]_{\lambda \mu}$ 
is a component of an irreducible tensor.  Then, by using 
\begin{equation}
\sum_{m_1 m_2} 
\left\langle l_1 l_2 m_1 m_2 \vert \lambda \mu\right\rangle
\ Y_{l_1 m_1}(\theta, \phi) Y_{l_2 m_2}(\theta, \phi)
= \sqrt{\frac{(2l_1 +1) (2l_2 + 1)}{4\pi (2\lambda + 1)}}
\left\langle l_1 l_2 0 0 \vert \lambda 0 \right\rangle
 Y_{\lambda \mu}(\theta, \phi)\ ,
\end{equation}
the second order term in Eq.~(\ref{Vib-fst}) can be written as
\begin{align}
T_2 &=\  
\frac{1}{2}
R_0^2\ \frac{\partial^2f_0(r)}{\partial r^2}\ 
\sum_{l_1 m_1 l_2 m_2 \lambda \mu K}
\left< l_1 l_2 m_1 m_2 \vert \lambda \mu \right>
\left[\alpha_{l_1}^\star \otimes \alpha_{l_2}^\star \right]_{\lambda
\mu}
\nonumber\\ 
&\hspace*{1.0cm}\times\sqrt{\frac{(2l_1+1)(2l_2+1)}{4\pi(2K+1)}}
\ \left< l_1 l_2 0 0 \vert K 0 \right>
\ \left< l_1 l_2 m_1 m_2 \vert K M_K \right>
\ Y_{KM_K}(\theta,\phi)\ .
\end{align}
The orthogonality of Clebsch-Gordan coefficients
reduces this to

\begin{equation}
T_2 =\ \frac{1}{2} R_0^2 
\ \frac{\partial^2f_0(r)}{\partial r^2}\ 
\sum_{\lambda}
\sqrt{\frac{(2l_1+1)(2l_2+1)}{4\pi(2\lambda + 1)}}
\ \left< l_1 l_2 0 0 \vert \lambda 0 \right>
\ \left[\alpha_{l_1}^\star \otimes \alpha_{l_2}^\star \right]_{\lambda}
{\bf \cdot  Y}_{\lambda}(\theta, \phi)\ ,
\label{T2-contract}
\end{equation}
since the generalised coefficients must satisfy the spherical
harmonic condition.

Then the function form can be recast as
\begin{align}
f(r) &= f_0(r) - R_0 \left(\frac{\partial f(r)}{\partial r}\right)_0 \sum_{\lambda} 
{\cal Q}^{(1)}_\lambda \cdot  Y_{\lambda}(\theta, \phi)
\nonumber\\
&\hspace*{5.0cm}+ \frac{1}{2} R_0^2 \left(\frac{\partial^2 f(r)}{\partial r^2} \right)_0
\sum_\lambda \left[\sum_{l_1 l_2} 
{\cal Q}^{(2)}_\lambda(l_1 l_2)\right] 
\cdot Y_{\lambda}(\theta, \phi) \ ,
\end{align}
where ${\cal Q}_\lambda^{(i)}$ are the first and (partial) second order Tamura
operators~\cite{Ta65},
\begin{equation}
{\cal Q}_{\lambda \mu}^{(1)} = \alpha_{\lambda \mu}^\star\ ;\hspace*{0.3cm}
{\cal Q}_{\lambda \mu}^{(2)}(l_1 l_2) = 
\sqrt{\frac{(2l_1+1)(2l_2+1)}{4\pi (2\lambda +1)}} 
\left\langle l_1 l_2 0 \vert 0 \lambda 0 \right\rangle
\left[\alpha_{l_1}^\star \otimes \alpha_{l_2}^\star \right]_{\lambda \mu}\ .
\end{equation}
\subsection{The nucleus as  a quantised liquid drop}

With the surface of a liquid drop of incompressible fluid that can be slightly 
deformed represented as in Eq.~(\ref{dropsur}), and with $\lambda \ge 2$,
quantization proceeds by mapping the generalised coordinates 
($\alpha_{\lambda \mu}$) and their canonical generalised momenta
using boson creation/annihilation operators ($b^\dagger_{\lambda \mu}/ 
b_{\lambda \mu}$), by
\begin{equation}
\alpha_{\lambda \mu} \Rightarrow \sqrt{\frac{\hbar}{2B_\lambda \omega_\lambda}}
\left[b_{\lambda \mu} + (-)^\mu b^\dagger_{\lambda -\mu} \right]\ .
\end{equation}
With a similar form for the generalised momentum, the Hamiltonian for a 
vibrating liquid (quantal) drop is
\begin{equation} 
H = \sum_{\lambda \mu} \left[ b^\dagger_{\lambda \mu} b_{\lambda \mu}  
+ \frac{1}{2} \right] \hbar \omega_\lambda\hspace*{0.5cm} 
{\rm where}\hspace*{0.5cm}
\left[b_{\lambda \mu}, b^\dagger_{\lambda^\prime \mu^\prime} 
\right] =
 \delta_{\lambda \lambda^\prime} \delta_{\mu \mu^\prime}\ .
\end{equation}
Then, normalized 1 and 2 phonon states are defined by
\begin{align}
&|1;\lambda \mu\rangle = b^\dagger_{\lambda \mu} |0\rangle\ ,
\nonumber\\
&|2;\left(\lambda_1 \lambda_2\right)  J  M\rangle = 
\frac{1}{\sqrt{1 + \delta_{\lambda_1 \lambda_2}}} 
\left[ b_{\lambda_1}^\dagger \otimes b_{\lambda_2}^\dagger 
\right]_{ J  M} |0\rangle\ ,
\label{12phonon}
\end{align}
where
\begin{equation}
\left[ b_{\lambda_1}^\dagger \otimes b_{\lambda_2}^\dagger 
\right]_{ J  M}
= \sum_{m_1 m_2} \left< \lambda_1 \lambda_2 \mu_1 \mu_2 | J M\right>
\ \  b_{\lambda_1 \mu_1}^\dagger  b_{\lambda_2 \mu_2}^\dagger\ . 
\end{equation}
This model involves generalised mass and restoring force parameters
($B_\lambda, C_\lambda$) with which the frequencies of the phonons and the 
coupling parameters are 
\begin{equation}
\omega_\lambda = \sqrt{\frac{C_\lambda}{B_\lambda}}\ ;\hspace*{0.5cm}
\beta_\lambda = \sqrt{(2\lambda + 1)} \left(a_\lambda \right)_0 
= \vert \left\langle 1 \left\| \alpha_\lambda \right\| 0\right\rangle\vert
= \sqrt{(2\lambda + 1)} \sqrt{\frac{\hbar}{2B_\lambda \omega_\lambda}}\ . 
\end{equation}
Here $\left(a_\lambda \right)_0$ is the zero point amplitude of 
vibration~\cite{Bo75,Gr96}.

\subsection{The vibration model for coupled-channels interactions}

With channels $c = \{lj;I;{ J}{ M} \}$ as used in the
MCAS theory of a nucleon interacting with a nucleus~\cite{Am03},
matrix elements of the type
\begin{align}
&\left[ f(r) \right]_{cc^\prime} = \left[ f_0(r) \right]_{cc^\prime}
- R_0 \left[\frac{\partial f(r)}{\partial r}
\sum_\lambda \left[\alpha_\lambda^\star \cdot Y_\lambda(\theta \phi)\right] 
\right]_{cc^\prime}\nonumber\\
&\; \; + \frac{1}{2} R_0^2 \left[ \left(\frac{\partial^2 f(r)}{\partial r^2} \right)_0 
\sum_\lambda \sum_{l_1 l_2} 
\sqrt{\frac{(2l_1+1)(2l_2+1)}{4\pi (2\lambda+1)}} 
\left\langle l_1 l_2 0 0 \vert \lambda 0 \right\rangle
\left[\alpha_{l_1}^\star \otimes \alpha_{l_2}^\star\right]_\lambda
\cdot Y_\lambda(\theta \phi) \right]_{cc^\prime}\ ,
\label{frexpand}
\end{align}
are required.  We choose to use as the basic interaction potential  form,
\begin{equation}
f_0(r) = 
\left[ V_0 + V_{ll} {\{\bf l \cdot l}\} + V_{ss} {\{\bf I \cdot s}\} 
\right] w(r) + 2\lambda_\pi^2 
V_{ls} \frac{1}{r} \frac{\partial w(r)}{\partial r} {\{\bf l \cdot s}\}\ .
\label{poteq}
\end{equation}
A Woods-Saxon form, $w(r) = \left[1 + \exp\left(\frac{r-R_0}{a} \right) \right]^{-1}$ is  used.

Each operator character of  the interaction  has
zero, first, and second order
elements due to the expansion in deformation.  Thus for each term in the
interaction, form factors in whatever channel coupling can be specified as
\begin{align}
V_{cc'}(r) = \left\{V^{(0)}(r)\right\}_{cc'} 
&+\;  \left\{V^{(1)}(r) \sum_\lambda 
{\mathbf {\cal Q}_\lambda^{(1)} \cdot Y_\lambda}(\theta \phi) \right\}_{cc'} 
\nonumber\\
&\hspace*{0.5cm}+\; \left\{V^{(2)}(r) \sum_\lambda 
\left[ \sum_{l_1 l_2} {\cal Q}_\lambda^{(2)}(l_1, l_2)\right] 
\cdot Y_\lambda(\theta \phi) \right\}_{cc'}\ .
\label{Mcaspot1}
\end{align}
With $W_{ls} = 2 \lambda_\pi^2 V_{ls}$, the zero order term in
Eq.~(\ref{Mcaspot1}) is
\begin{align}
\left\{ V^{(0)}(r) \right\}_{c c^\prime} =& \left\{ 
\left[ V_0^{(c)} + V_{ll}^{(c)} l(l+1) 
\right] w(r) + W_{ls}^{(c)} \frac{1}{r} \frac{\partial w(r)}{\partial r} {\{\bf l \cdot s}\}
\right\} \delta_{c c^\prime}
\nonumber\\
&\hspace*{4.0cm}
+ \frac{1}{2} \left[ V_{ss}^{(c)} + V_{ss}^{(c^\prime)} \right] w(r)
\  {\{\bf I \cdot s}\}_{cc^\prime} \ ,
\label{Vcc-zero}
\end{align}
and the superscripts on the potential strengths indicate that the
values for the appropriate parities of the channel $c$ are to be taken.
Also, a symmetrized form is used for terms that allow coupling between
different channels. Such is the case with the other two components
and
\begin{align}
& \left\{ V^{(1)}(r)\right\}_{cc^\prime} = \left\{
- R_0 \frac{\partial w(r)}{\partial r} \
\ \frac{1}{2} \left[ V_0^{(c)} + V_0^{(c^\prime)} 
+  V_{ll}^{(c)} \left\{{\bf l\cdot l}\right\}_{cc}
+ V_{ll}^{(c^\prime)}
\left\{{\bf l\cdot l}\right\}_{c^\prime c^\prime}
\right] \right.
\nonumber\\
&\hspace*{3.5cm} 
- \left.
\frac{1}{2}\ R^2_0\ \frac{1}{r}\ 
\frac{\partial^2 w(r)}{\partial r^2}  
\Bigl( W_{ls}^{(c)} \left\{{\bf l\cdot s}\right\}_{cc}
+ W_{ls}^{(c^\prime)} \left\{{\bf l\cdot s}\right\}_{c^\prime c^\prime}\Bigr)
\right\}
\sum_L \left[{\cal Q}_L^{(1)} \cdot Y_L \right]_{cc^\prime}
\nonumber\\
&\hspace*{1.2cm}  
- \frac{1}{2}\ R_0\ \frac{\partial w(r)}{\partial r} \sum_L \sum_{c^{\prime \prime}} 
\left\{ V_{ss}^{(c^\prime)}
\left[{\cal Q}_L^{(1)} \cdot Y_L \right]_{cc^{\prime\prime}}
\left[{\bf I \cdot s}\right]_{c^{\prime \prime}c^\prime}
+ V_{ss}^{(c)}
\left[{\bf I \cdot s}\right]_{c c^{\prime \prime}}
\left[{\cal Q}_L^{(1)} \cdot Y_L \right]_{c^{\prime \prime}c^\prime}
\right\} .
\label{Vcc-one}
\end{align}
The second order terms are
\begin{align}
\left\{ V^{(2)}(r)\right\}_{cc^\prime} =& \left\{
\frac{1}{4} R_0^2\ \frac{\partial^2 w(r)}{\partial r^2} \left[ 
V_0^{(c)} + V_0^{(c^\prime)} 
+ V_{ll}^{(c)} \left\{{\bf l\cdot l}\right\}_{cc}
+
 V_{ll}^{(c^\prime)} \left\{{\bf l\cdot l}\right\}_{c^\prime c^\prime}
\right] \right.
\nonumber\\
+ \frac{1}{4}\ R_0^3 & \left. \frac{1}{r}\ 
\frac{\partial^3 w(r)}{\partial r^3}  
\Bigl(W_{ls}^{(c)}\ \left\{{\bf l\cdot s}\right\}_{cc}
+ W_{ls}^{(c^\prime)}\ \left\{{\bf l\cdot s}\right\}_{c^\prime c^\prime} 
\Bigr) \right\}
\sum_\lambda \left[ \left\{\sum_{l_1 l_2} {\cal Q}_\lambda^{(2)}(l_1, l_2)
\right\} \cdot Y_\lambda \right]_{cc^\prime}
\nonumber\\
+ \frac{1}{4}\ R_0^2 & \frac{\partial^2 w(r)}{\partial r^2} 
\sum_\lambda
\sum_{c^{\prime \prime}} \left[ V_{ss}^{(c)}\ 
\left\{{\bf I\cdot s}\right\}_{cc^{\prime \prime}} 
\left[
\left\{\sum_{l_1 l_2} {\cal Q}_\lambda^{(2)}(l_1, l_2)\right\} 
\cdot Y_\lambda \right]_{c^{\prime \prime}c^\prime}
\right.\nonumber\\
&\hspace*{5.0cm}\left.
+ V_{ss}^{(c^\prime)}\
\left[
\left\{\sum_{l_1 l_2} {\cal Q}_\lambda^{(2)}(l_1, l_2)\right\} 
\cdot Y_\lambda \right]_{c c^{\prime \prime}}
\left\{{\bf I\cdot s}\right\}_{c^{\prime \prime} c^\prime} 
\right]\ .
\label{Vcc-two}
\end{align}

The matrix elements of the operators 
$\left\{\bf l \cdot l\right\}$, $\left\{\bf I \cdot s \right\}$, and 
$\left\{\bf l \cdot s\right\}$ have been defined previously~\cite{Am03}.
And, as the first and second order terms require development as matrix elements
of nuclear operators, we use the Edmond's form of the Wigner-Eckart 
theorem, i.e.
\begin{equation}
\left\langle J_f M_f \left| T_{LM} \right | J_i M_i \right\rangle  = 
\frac{1}{\sqrt{(2J_f + 1)}} 
\left\langle J_i L M_i M \vert J_f M_f \right\rangle
\left\langle J_f \left|\right| T_L \left|\right| J_i\right\rangle \ .
\label{Edmond-WE}
\end{equation}
For the case of scalar operators to be used herein (so conserving
total angular momentum ${ J} = { J}^\prime$), specifically
with the Tamura operators cast temporarily as a general operator 
${\cal Q}$, we use
$T_{0,0} = \left[ {\cal Q}_L \cdot Y_L \right]_{0,0}$, so that
\begin{equation}
\left\langle c \left|\left[ {\cal Q}_L \cdot Y_L \right]_{0,0}
\right| c^\prime \right\rangle
=\, \frac{1}{\sqrt{(2{ J} + 1)}}\,
\left\langle c \left|\right| \left[ {\cal Q}_L \cdot Y_L \right]_0
\left| \right| c^\prime\right\rangle \ ,
\end{equation}
for all ${ M}$ as the Clebsch-Gordan coefficient is a delta function.  
Then using a Brink and Satchler identity (Eq.~(5.13) in ~\cite{Br68}), 
suitably adjusted to Edmond's form for the Wigner-Eckart theorem, i.e.
\begin{align}
\left\langle c \left|\right| \left[{\cal Q}_L \cdot Y_L \right]
\left| \right| c^\prime \right\rangle &=
\left\langle (jI){ J} \left|\right| \left[ {\cal Q}_L \cdot Y_L \right]
\left|\right| (j^\prime I^\prime){ J} \right\rangle
\nonumber\\
&=
(-)^{j^\prime + I + { J}} 
\left\{\begin{array}{ccc}
j & j^\prime & L \\ 
I^\prime & I & { J}
\end{array}
\right\}
\frac{1}{\sqrt{(2{ J}+1)}} 
\left\langle j \left|\right| Y_L \left|\right| j^\prime \right\rangle
\left\langle I \left|\right| {\cal Q}_L \left|\right| I^\prime \right\rangle \ ,
\end{align}
and as
\begin{equation}
\left\langle j \left|\right| Y_L \left|\right| j^\prime \right\rangle 
= (-)^{j + L - j^\prime} 
\sqrt{\frac{(2L+1)(2j^\prime+1)}{4\pi}}
\left\langle j^\prime  L  \sfrac{1}{2}  0 \bigg| j  \sfrac{1}{2} \right\rangle \ ,
\end{equation}
\begin{eqnarray}
&&
\left\langle c \left| \right|  \left[ {\cal Q}_L \cdot Y_L(\Omega) \right] 
\left|\right| c^\prime  \right\rangle =
(-)^{j + L + I + { J}} 
\left\{ \begin{array}{ccc}
j & j^\prime & L \\
 I^\prime & I  & { J}
\end{array}
\right\}
\sqrt{\frac{(2j^\prime+1_)(2L+1)} {4\pi (2{ J}+1)}}
\nonumber\\
&&\hspace*{6.5cm}\times 
\left\langle j^\prime  L  \sfrac{1}{2}  0 \bigg|  j  \sfrac{1}{2}\right\rangle
\left\langle I \left|\right| {\cal Q}_L \left|\right| I^\prime \right\rangle\ . 
\label{rme-vibmod}
\end{eqnarray}
Thus to specify all terms in the form for the interaction matrix of
potentials, Eq.~(\ref{Vcc-two}), the reduced matrix elements of the 
various Tamura operators~\cite{Ta65} are required.

\begin{acknowledgments}
SK acknowledges support from the National Research Foundation of South Africa. JPS acknowledges grant support from NSERC, Canada, during the early part of this work.
\end{acknowledgments}

\bibliography{Nucleon-16O-rev7b-2016.bib}

\end{document}